\pdfoutput=1

\documentclass[11pt,letter,subeqn,fleqn]{article}
\usepackage{amsmath, amssymb,amsthm,cite}
\usepackage[normal]{subfigure}
\usepackage{subfig}
\usepackage{graphicx}
\usepackage{tabulary}
\usepackage{float}
\usepackage[toc,page]{appendix}

\title{Nonlinear Dynamics of a Viscous Bubbly Fluid}

\author{ 
Ryan J. Thiessen\footnotemark[1], \quad Alexei F. Cheviakov\footnotemark[2]\vspace{0.5cm}\\
\small \emph{Department of Mathematics and Statistics, University of Saskatchewan, Saskatoon, Canada}\vspace{0.2cm}
 }

 \small\normalsize

\def\beq{\begin{equation}}
\def\eeq{\end{equation}}
\def\barr{\begin{array}{ll}}
\def\earr{\end{array}}

\def\const{\hbox{\rm const}}

\def\grad{\mathop{\hbox{\rm grad}}}

\def\div{{\hbox{\rm div}}}
\def\grad{{\mathop{\hbox{\rm grad}}}}

\def\curl{\mathop{\hbox{\rm curl}}}

\def\vec#1{{\boldsymbol{\rm #1}}} 

\def\Re{{\rm Re}}
\def\Eu{{\rm Eu}}

\def\B{{B}}
\def\const{\hbox{\rm const}}

\def\der#1#2{\dfrac{d #1}{d #2}}
\def\pder#1#2{\dfrac{\partial #1}{\partial #2}}
\def\pderr#1#2{\dfrac{\partial^2 #1}{\partial {#2}^2}}
\def\pderflat#1#2{{\partial #1}/{\partial #2}}

\def\matD{\mathcal{D}}

\newtheorem{proposition}{Proposition}

{\theoremstyle{definition} 

\newtheorem{remark}{Remark}
}

\newcounter{tabnum}

\renewcommand{\div}{\mathrm{div}}

\begin{document}


\footnotetext[1]{Corresponding author. Electronic mail: rjt128@mail.usask.ca}
\footnotetext[2]{Alternative English spelling: Alexey Shevyakov. Electronic mail: shevyakov@math.usask.ca}

\maketitle \numberwithin{equation}{section}
\maketitle \numberwithin{remark}{section}
\numberwithin{lemma}{section}
\numberwithin{proposition}{section}

\begin{abstract}
A physical model of a three-dimensional flow of a viscous bubbly fluid in an intermediate regime between bubble formation and breakage is presented. The model is based on mechanics and thermodynamics of a single bubble coupled to the dynamics of a viscous fluid as a whole, and takes into account multiple physical effects, including gravity, viscosity, and surface tension. Dimensionless versions of the resulting nonlinear model are obtained, and  values of dimensionless parameters are estimated for typical magma flows in horizontal subaerial lava fields and vertical volcanic conduits.

Exact solutions of the resulting system of nonlinear equations corresponding to equilibrium flows and traveling waves are analyzed in the one-dimensional setting. Generalized Su-Gardner-type perturbation analysis is employed to study approximate solutions of the model in the long-wave ansatz. Simplified nonlinear partial differential equations (PDE) satisfied by the leading terms of the perturbation solutions are systematically derived. It is shown that for specific classes of perturbations, approximate solutions of the bubbly fluid model arise from solutions of the classical diffusion, Burgers, variable-coefficient Burgers, and Korteweg-de Vries equations.


\end{abstract}

\section{Introduction}

The derivation of physically sound models describing the dynamics of multiphase flows in various settings is an important topic of active research. Multiphase flows arise in a vast variety of physical and engineering applications; they can involve single or multiple fluid, solid and gas phases, phase interfaces, surfactant dynamics, porous media, hard and soft particles, gas bubbles, droplets, anisotropy effects, and multiple other phenomena and aspects (see, e.g., Refs.~ \cite{nigmatulin1990dynamics, gidaspow1994multiphase, kolev2005multiphase, brennen2005fundamentals, passman1984theory} and references therein). The description of such flows, including the formulation of constitutive relations, and the analysis and solution of the resulting models, including numerical simulation, present significant challenges. Vast literature is devoted to multiple aspects of multiphase dynamics; some examples of recent works dedicated to general analysis and specific applications are found in Refs.~\cite{melnik, Kud_Sine, wang2011thermodynamic, kallendorf2012conservation, cheviakov2017symbolic, savage1979gravity}.

Multiphase flows are used in mathematical description of many geological processes related to volcanic activity, including, for example, pyroclastic flows consisting of volcanic gas and solid tephra particles, or magma/lava flows containing gas bubbles \cite{melnik, Kud_Sine}. Due to their enormous viscosities and high temperatures, basaltic lava sheet flows provide an extreme example of multiphase flows \cite{hon1994emplacement, JGRB:JGRB11445, sakimoto1998flow, park1984dynamics}. These sheet flows are characterized by an uninterrupted motion of lava, which produces a smooth upper and lower crust that acts as a insulator. Most significant difficulties in modeling such flows arise in the conjoining of the microscopic and macroscopic scales present in the system. The microscopic scale corresponds to a single bubble and usually involves analysis of two phases separated by an interface. The first to study this was Rayleigh \cite{rayleigh1917viii}, describing the collapse of a empty spherical cavity in a fluid. Later Plesset \cite{plesset1954growth} generalized Rayleigh's setup to include the gas inside the sphere and viscosity of the surrounding fluid. Studying the dynamics of the bubbles surface, Plesset found that the bubbles radius $R = R(t)$ satisfied a single second-order ordinary differential equation (ODE), termed the \emph{Rayleigh-Plesset equation} (see Section \ref{physModel} below). The macroscopic scale of the problem corresponds to the mixture flow, and commonly modeled as a homogenous continuum governed by Euler or Navier-Stokes fluid dynamics equations. In this context, it is important to mention the works of Foldy \cite{foldy1945multiple} and  Wijngaarden \cite{wijngaarden1972one} (see also Ref.~\cite{miksis1991effective}). In particular, Wijngaarden showed that  the propagation of linear one-dimensional waves in an isothermal bubbly fluid was governed described by a linear 4th-order partial differential equation (PDE). Solutions to this equation were later studied by Jordan and Feuillade \cite{jordan2004propagation, jordan2006propagation}, who also noted that their numerical solution behaved qualitatively similarly to solutions of the Burgers equation. Moving on from acoustic waves, Wijngaarden was also interested in other applications of bubbly fluids, where the nonlinear effects would be larger. For example, bubbly liquids in the field of ultrasonics where gas micro-bubbles are used as contrast agents to enhance the acoustic contrast between blood and surrounding tissues \cite{becher2012handbook, goldberg1994ultrasound, szabo2004diagnostic}. These problems have recently been formalised extended to include nonlinear waves \cite{kanagawa2015two,kanagawa2011nonlinear,kanagawa2011nonlinear}. Considering a small perturbation in the pressure, it was shown that the Euler fluid equations in one dimension and the Rayleigh-Plesset equation reduce to a nonlinear Burgers-KdV equation \cite{wijngaarden1972one}. This is in agreement with the results of the work of Nakoryakov, Sobolev and Shreiber \cite{Nakoryakov1972}, where long wavelength perturbations of the same model were analyzed. A related recent work of Kudryashov and Sinelshchikov \cite{Kud_Sine} which motivated the research presented in this paper is devoted to macroscopic modeling and analysis of a bubbly fluid; it investigates the Rayleigh-Plesset equation coupled with the Euler fluid dynamics equations and an additional inter-phase heat transfer equation.  In the long-wavelength limit, the leading term of the asymptotic series for the solution of the model was claimed to satisfy an interesting third-order nonlinear PDE (in fact, a family of PDEs involving parameters), generalizing Burgers, Korteweg-de Vries, and Burgers-Korteweg-de Vries equations. This PDE, later named the \emph{Kudryashov-Sinelshchikov equation}, has attracted significant attention on its own right as a new equation of mathematical physics. Kudryashov and Sinelshchikov further extended their model in Ref.~\cite{kudryashov2014extended}, where small effects of liquid compressibility and
surface tension were considered, leading to third- and fourth- order partial differential equations. In parallel, Kanagawa \emph{et al} \cite{kanagawa2010unified} proposed a systematic derivation of nonlinear equations in bubbly liquids, based on parameter scaling. Their starting point was similar to that of the work of Kudryashov and Sinelshchikov, but Keller's equation was used to describe the oscillations of bubble; this led to models involving Burgers-Korteweg-de Vries and nonlinear Schr\"{o}dinger equations.

The goal of the current contribution is the formulation of a more general model of a viscous bubbly flow, taking into account a wider set of physical effects compared to Refs.~\cite{Kud_Sine,kudryashov2014extended,kanagawa2010unified}. In particular, the new model satisfies the important physical requirement of Galilean invariance; it is based on Navier-Stokes equations in order to fully treat viscosity effects; it includes gravity, bubble surface tension, and a physically sound heat flux relationship. Typical applications for the presented model can range from subaerial lava flow fields to industrial oils carrying bubbles. We then undertake some analysis of the model, deriving its dimensionless versions, estimating typical parameter values, and seeking its exact and approximate solutions.

The paper is organized as follows. Section \ref{physModel} is dedicated to the derivation of the bubbly fluid model, which includes the consideration of mechanics and thermodynamics of a single gas bubble and the bubbly fluid as a whole. Two dimensionless versions of the governing equations are derived. First, a general non-dimensionalization is presented, aiming at the maximal reduction of the number of constants in the model, contains only three dimensionless physical parameters, as compared to seven in the dimensional equations. A ``physical" non-dimensionalization is also presented, which uses characteristic values of bubbly fluid parameters instead; it contains the well-known dimensionless quantities such as Reynolds and Euler numbers, the ratio of characteristic spatial dimensions, etc. Equilibrium and traveling wave solutions of the bubbly fluid model in one spatial dimension are also considered in this section.

Section \ref{pertAnalysis} is devoted to the analysis of asymptotic expansions of solutions of the bubbly fluid model in one spatial dimension about a constant equilibrium state, using the Su-Gardner-type generalized power series approach, and a Gardner-Morikawa rescaled coordinates in a moving frame of reference, designed  to capture long-wavelength, slow-time perturbations. For specific classes of the scale transformation parameters of the asymptotic solutions, we show that the leading non-constant terms of the perturbation series may satisfy the linear diffusion equation, or the nonlinear Burgers or Korteweg-de Vries equations. 

In Section \ref{sec:grav}, we generalize the asymptotic analysis of Section \ref{pertAnalysis} to cases of flows with a nonzero gravity term. We show that it is also possible to reduce the problem of finding the lowest-order perturbation terms to a solution of a single PDE. Interestingly, for gravity-driven flows, this PDE is a variable coefficient Burgers-type equation, and the frame of reference must be traveling with a variable speed, prescribed by the parameters of the process.


In Section \ref{sec:surfT}, the bubble flow model is further extended by considering an extra term in the Rayleigh-Plesset equation corresponding to the bubble surface tension. It is shown that leading-order terms of the asymptotic perturbations of equilibria for such flows in one spatial dimension can also be described by a single PDE, such as linear diffusion, Burgers, or Korteweg-de Vries equation. 

The important fact that for a set of rather complex nonlinear models (the bubble-liquid mixture flow model with and without gravity and surface tension), the leading perturbation series solution terms arise from a solution of a classical (diffusion, Burgers, or Korteweg-de Vries) equation, allows one to use known exact closed-form solutions of these PDEs to produce physically meaningful approximate closed-form solutions of the full nonlinear model. This idea is illustrated with a computational example of an approximate traveling kink-type solution in Section \ref{sec:ND:Burg}. 

The paper is concluded with an overview of the results and their discussion in Section \ref{sec:Conclusion}.

\section{The physical model of a one-dimensional bubbly fluid flow}\label{physModel}

We develop a model of dynamics of a viscous liquid with gas bubbles, describing one-dimensional vertical and horizontal flows, and taking into account the heat transfer between the gas and liquid phases. In the derivation, relevant results from Refs.~\cite{nakoryakov1993wave, Kud_Sine} are employed.




\subsection{The derivation of the PDE model}

We use subscripts ``1" and ``2" to refer to the fluid and gas phases respectively (see Figure \ref{fig:bubb}). The variables and parameters with no subscript refer to the corresponding quantities for the mixture of the liquid and gas bubbles. In particular, the densities of the fluid and gas phases are given by $\rho_1$ and $\rho_2$, the temperatures by $T_1$ and $T_2$, the gas pressure is denoted $P_2$, whereas the pressure of the fluid and the mixture are identified: $P_1=P$. Other parameters are denoted accordingly. The all the physical parameters of the mixture in the final model are functions of the time $t$ and the spatial variable $\vec{x}=(x,y,z)\in \mathbb{R}^3$.


The model presented below describes viscous bubbly fluids in an intermediate regime between bubble formation and breakage, i.e., when bubbles are small and well-separated. In particular, the following physical assumptions are used below: the density and viscosity of the fluid phase are constant; bubbles do not form or disappear; bubbles and fluid do not exchange mass; the effect of bubble surface tension is negligible (the latter requirement is removed in Section \ref{sec:surfT} below). The exchange of mass between the bubbles and the fluid is small, for example, in certain regimes of magma flows within the volcanic conduit and subaerial flow fields.


\subsubsection{Mechanics of a single gas bubble} \label{sec:bubble:mech}

In order to describe the dynamics of parameters of well-separated bubbles within the mixture, consider a single isolated spherical bubble of radius $R = R(t)$. Consequently, $r\geq R(t)$ corresponds to the liquid phase, and $r < R(t)$ to the gaseous phase, where $r$ is the spherical radial variable (Figure \ref{fig:bubb}).
\begin{figure}[H]
\centering
\includegraphics[width=0.6\textwidth]{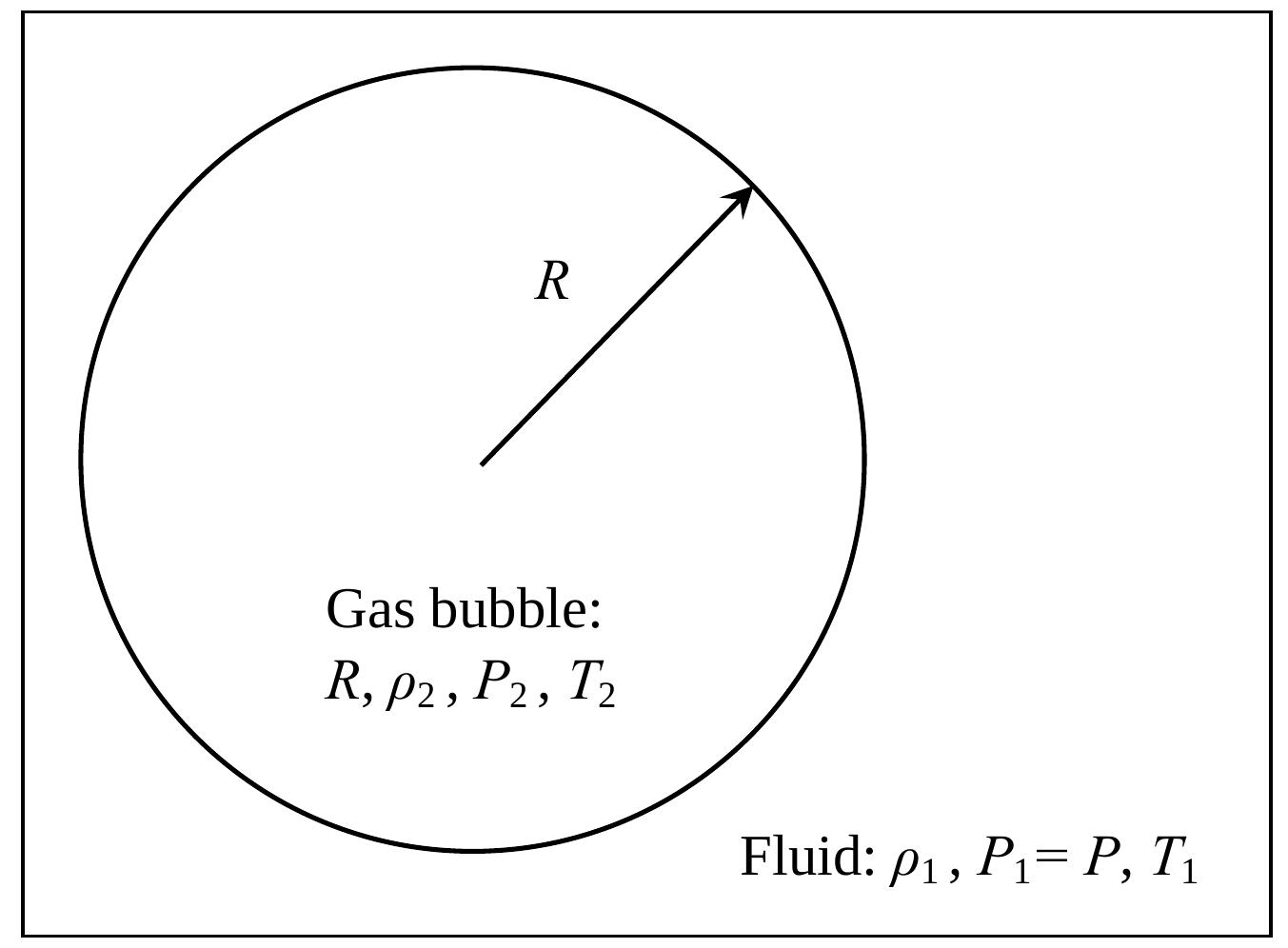}
\caption{\label{fig:bubb}An isolated spherical bubble.}
\end{figure}

The equations of motion for the incompressible fluid outside the bubble given by the Navier-Stokes equations:
\begin{equation} \label{u1momcom}
\pder{\vec{u}_1}{t} + (\vec{u}_1\cdot\nabla)\vec{u}_1 + \frac{1}{\rho_1}\grad \,P_1  =  \frac{\mu}{\rho_1} \Delta \vec{u}_1,
\end{equation}
\begin{equation} \label{imcomp}
\div\,\vec{u}_1 = 0.
\end{equation}
Which can be reduced to the \emph{Rayleigh-Plesset equation} (see \cite{leighton2007derivation,brennen2013cavitation} for details)
\begin{equation} \label{eq:RPle}
P = P_2 -  \rho_1\left(R\der{^2 R}{t^2} + \frac{3}{2}\left(\der{R}{t}\right)^2 + 	\frac{4{\mu}}{3\rho_1 R}\der{R}{t}\right),
\end{equation}
relating the bubble radius $R(t)$, the gas pressure $P_2(t)$ within the bubble, and the pressure mixture $P(t)$ away from the bubble.

\subsubsection{Thermodynamics of a gas bubble} \label{sec:bubble:therm}

An additional connection between the gas pressure and bubble radius follows from thermodynamical relationships, in particular, the energy balance equation, which in three dimensions has the form
\begin{equation} \label{eq:gas:en}
\rho_2 \left(\pder{ e_2}{t} + \vec{u}_2\cdot \grad\, e_2\right)=  - \div\, \vec{q} - P_2\,\div\,\vec{u}_2,
\end{equation}
where $e_2=C_{v2}T_2$ is the thermal energy density (per unit mass) of the gas phase,  $C_{v2}=\const$ is the specific heat of the gas phase at the constant volume, and $\vec{q}$ is the energy flux vector.
The two additional equations one needs to take into account is the gas momentum balance  given by
\begin{equation} \label{eq:gas:mom}
\rho_2 \left(\der{\vec{u}_2}{t} + (\vec{u}_2\cdot\nabla)\vec{u}_2\right)   = -\grad \,P_2,
\end{equation}
where the gas viscosity was neglected, and the continuity equation
\begin{equation} \label{eq:gas:mass}
\der{\rho_2}{t}  + \div\,(\rho_2 \vec{u}_2)=0,
\end{equation}
The above system is closed with the ideal gas equation
\begin{equation} \label{eq:gas:idgas}
P_2=(\gamma-1)C_{v2}\rho_2 T_2,
\end{equation}
with $\gamma=\const$ denoting the adiabatic exponent, and $T_2$ the gas temperature.

In spherical coordinates, under the assumption of spherical symmetry ($\vec{u}_2 = u_2(r,t) \vec{e}_r$, and the dependence of all other gas parameters on $(r,t)$ only), the equations \eqref{eq:gas:en}, \eqref{eq:gas:mom}, and \eqref{eq:gas:mass} become respectively
\begin{subequations}
\begin{equation} \label{eq:gas:en:sph}
C_{v2} \rho_2 \left(\pder{T_2}{t} + {u}_2\pder{T_2}{r}\right)=  - \div\, \vec{q} - \dfrac{P_2}{r^2}\pder{}{r}\left(r^2 {u}_2\right),
\end{equation}
\begin{equation} \label{eq:gas:mom:sph}
\rho_2 \left(\pder{u_2}{t} + {u}_2\pder{u_2}{r}\right)=  -\pder{P_2}{r},
\end{equation}
\begin{equation} \label{eq:gas:mass:sph}
\pder{\rho_2}{t} + \dfrac{1}{r^2}\pder{}{r}\left(r^2\rho_2{u}_2\right)= 0.
\end{equation}
\end{subequations}
We solve \eqref{eq:gas:idgas} for $\rho_2$ and substitute the result into the remaining equations. Similarly, we solve \eqref{eq:gas:mom:sph} for $\pderflat{u_2}{t}$ and substitute the result into the remaining equations. Finally, the consequences of PDEs   \eqref{eq:gas:en:sph} and \eqref{eq:gas:mass:sph} both contain $\pderflat{T_2}{t}$. Excluding it, we arrive at the PDE
\begin{equation} \label{eq:gas:sph:noTt}
\pder{P_2}{t} =- \dfrac{\gamma P_2}{r^2}\pder{}{r}\left(r^2 {u}_2\right) - u_2\pder{P_2}{r}  - (\gamma-1)\div\, \vec{q}.
\end{equation}
Under an additional physical assumption of fast pressure re-balancing inside the bubble, $\pderflat{P_2}{r}=0$, the equation \eqref{eq:gas:sph:noTt} may be completely integrated in $r$ from $0$ to $R(t)$. The boundary conditions to be used are given by
\[
u_2\Big|_{r = 0} =0,\qquad u_2\Big|_{r = R(t)}=\frac{d R(t)}{d t}.
\]
Moreover, due to the spherical symmetry, $\vec{q} = q(r,t) \vec{e}_r$, and through the use of the divergence theorem, one has
\[
\int_{0}^{R(t)} r^2 \div\, \vec{q}\, dr = \dfrac{1}{4\pi} \int_{B}\div\, \vec{q}\, dV = q(t,R(t))R^2(t).
\]
where $B$ denotes the spherical bubble domain of radius $R(t)$. With these ingredients, the integrated version of \eqref{eq:gas:sph:noTt}  yields and ODE
\begin{equation} \label{heatTransfer0}
\frac{d P_2}{d t}  + \frac{3\gamma P_2}{R}\frac{d R}{d t} = - \frac{3(\gamma - 1)}{R}\, q(t,R),
\end{equation}
with $P_2=P_2(t)$, $R=R(t)$.

In order to use the ODE \eqref{heatTransfer0}, a physical constitutive assumption regarding the form of $q(t,R)$ is required. For example, in \cite{Kud_Sine}, the constant value of Nusselt number was assumed. It is not clear, nor has an explanation been offered, why that condition might  physically hold. A natural choice we make here would be a \emph{generalized Newton's law of cooling}, stating that the outside-directed total heat flux through the boundary of the spherical bubble a domain is proportional to a power of the difference of temperatures immediately inside and outside the bubble:
\begin{equation} \label{eq:Newton}
q(R,t)=  h\left(T_2(R,t) - T_1(R,t)\right)^k \Big|_{r = R(t)}.
\end{equation}
Here $k>0$ is some constant power, and $h$ is the heat transfer coefficient, also assumed constant. In this work, we take the simplest case of $k=1$, corresponding to the classical Newton's law of cooling. The heat balance equation for a bubble becomes
\begin{equation} \label{heatTransfer1}
\frac{d P_2}{d t}  + \frac{3\gamma P_2}{R}\frac{d R}{d t} = - \frac{3(\gamma - 1)h}{R} \left(T_2(R,t)-T_1(R,t) \right).
\end{equation}
Suppose that initially, both the gas in the bubble and the surrounding fluid had he same temperature $T_0$. For fluids with large specific heats, it is natural to assume $T_1(R,t)=T_0=\const$. We also assume $T_2=T_2(t)$ (the temperature on the bubble surface is close to the average temperature throughout the bubble). The equation of state for the ideal gas in the bubble as a whole reads
\[
P_2(t) V_2(t)=n R^* T_2(t),
\]
where $R^*$ is the ideal gas constant, $n$ is the constant amount of substance in the bubble, and $V_2(t)={4}/{3}\pi R^3(t)$ is the bubble volume. Writing this formula at $t=0$ and at an arbitrary time $t$, and dividing, we obtain
\begin{equation}\label{eq31}
T_2(t)=\dfrac{P_2(t) T_{0}}{P_{0}}\frac{R^3(t)}{{R_0}^3},
\end{equation}
where the initial conditions are given by
\begin{equation}\label{eq:ICs}
P_2\big|_{t = 0}=P_{0},\qquad T_2\big|_{t = 0}=T_{0},\qquad R\big|_{t = 0}=R_0.
\end{equation}

The final result is an ODE relating the pressure $P_2(t)$, the radius $R(t)$, and the temperature $T_2(t)$ of the gas bubble. It is given by
\begin{equation} \label{heatTransfer3}
\frac{d P_2}{d t}  + \frac{3\gamma P_2}{R}\frac{d R}{d t} + 3(\gamma-1)\frac{{h}T_{0}}{R}{\left(\frac{P_2}{P_{0}}\frac{R^3}{{R_0}^3} - 1\right)} = 0 .
\end{equation}

\begin{remark}
If the gas within a bubble is assumed to undergo an adiabatic process, satisfying $PV^{\gamma} = \const$, one would have, for the whole bubble,
\begin{equation} \label{eq:adia}
{P_2}{R^{{3}{\gamma}}} = \const = {P_0}{R_0^{{3}{\gamma}}}.
\end{equation}
It is straightforward to verify that under \eqref{eq:adia}, the left-hand side of the heat transfer equation \eqref{heatTransfer0} vanishes; it follows that the ODE \eqref{heatTransfer3} is a direct generalization of the adiabatic law onto the case of nonzero  energy flux through the bubble boundary.
\end{remark}

\subsubsection{Mass balance of the gas-fluid mixture} \label{sec:mix:mass:bal}

The next local equation of the model is derived from the assumption that the total mass of the mixture  is conserved, and that the ratio between the masses of the liquid and the gas is constant, i.e., bubbles do not form or disappear. Consider a mixture containing $N$ gas bubbles per kilogram ($N =\const$). Let $\hat{V}$ denote the volume of the gas bubbles per kilogram of the mixture:
\begin{equation}\label{eq15}
\hat{V}=\frac{4}{3}\pi N R^3.
\end{equation}

The masses of the fluid and the gas in the bubbles, contained in one kilogram of the mixture,  are given by dimensionless quantities
\begin{equation}
\delta m_2\equiv X= \rho_2 \hat{V}, \qquad \delta  m_1=1- \delta m_2.
\end{equation}
These may be called the ``relative mass contents" of the two phases in the mixture.
The total mass of the mixture is given by $m=\rho V$ $[kg]$, where $V$ is the volume occupied by the mixture, and $\rho$ is the average density of the mixture.

The total volumes occupied by the fluid phase and the gas bubbles are respectively given by
\begin{equation}\label{eq17}
V_2=m\hat{V}=\rho V \hat{V}, \qquad V_1=V-V_1= V-\rho V \hat{V}.
\end{equation}
The mass of the mixture can be further written as
\begin{equation}\label{eq18}
m=\rho V=\rho_1 V_1 + \rho_2 V_2.
\end{equation}
Using \eqref{eq17}, one obtains the following relation \cite{Kud_Sine}:
\begin{equation}\label{eq19}
\left(1 - \rho_2\hat{V} + \rho_1 \hat{V}\right)\rho - \rho_1 = 0.
\end{equation}
One consequently has
\begin{equation}\label{gasliquidCons}
\left(1 - X + \frac{4}{3}{\pi}{N}R^3\rho_1\right)\rho - \rho_1 = 0,
\end{equation}
an algebraic formula connecting the total mixture density $\rho$, the density of gas in the bubbles  $\rho_2$, and the bubble radius $R$. In \eqref{gasliquidCons}, $N, X=\const$; moreover, the fluid density $\rho_1$ is also assumed constant for the purposes of the current model.

\subsubsection{Dynamics of the bubbly fluid} \label{sec:mix:dyn}

In order to describe the dynamics of the bubbly fluid as a whole, we step away from the previous description at the bubble scale, and use the one-dimensional Navier-Stokes equations. These consist of the continuity and the momentum equation, given by
\begin{equation} \label{massconservation}
\pder{\rho}{t} + \div(\rho \vec{u}) = 0,
\end{equation}
\begin{equation}\label{Momcon}
\rho \left(\pder{\vec{u}}{t} + (\vec{u}\cdot\nabla)\vec{u}\right) + \grad \,P =\mu \Delta{\vec{u}} - \rho g\vec{k},
\end{equation}
where $\rho  =\rho(\vec{x},t)$ is the density of the mixture, $u =u(\vec{x},t)$  is the velocity of the mixture in the $x$-direction, $P = P(\vec{x},t)$ is the pressure of the mixture, $g$ is the downward acceleration due to gravity, and $\vec{k}$ is the unit vector in the $z$-direction (Figure \ref{fig:bubb}).

\begin{figure}[H]
\centering
\includegraphics[width=0.6\textwidth]{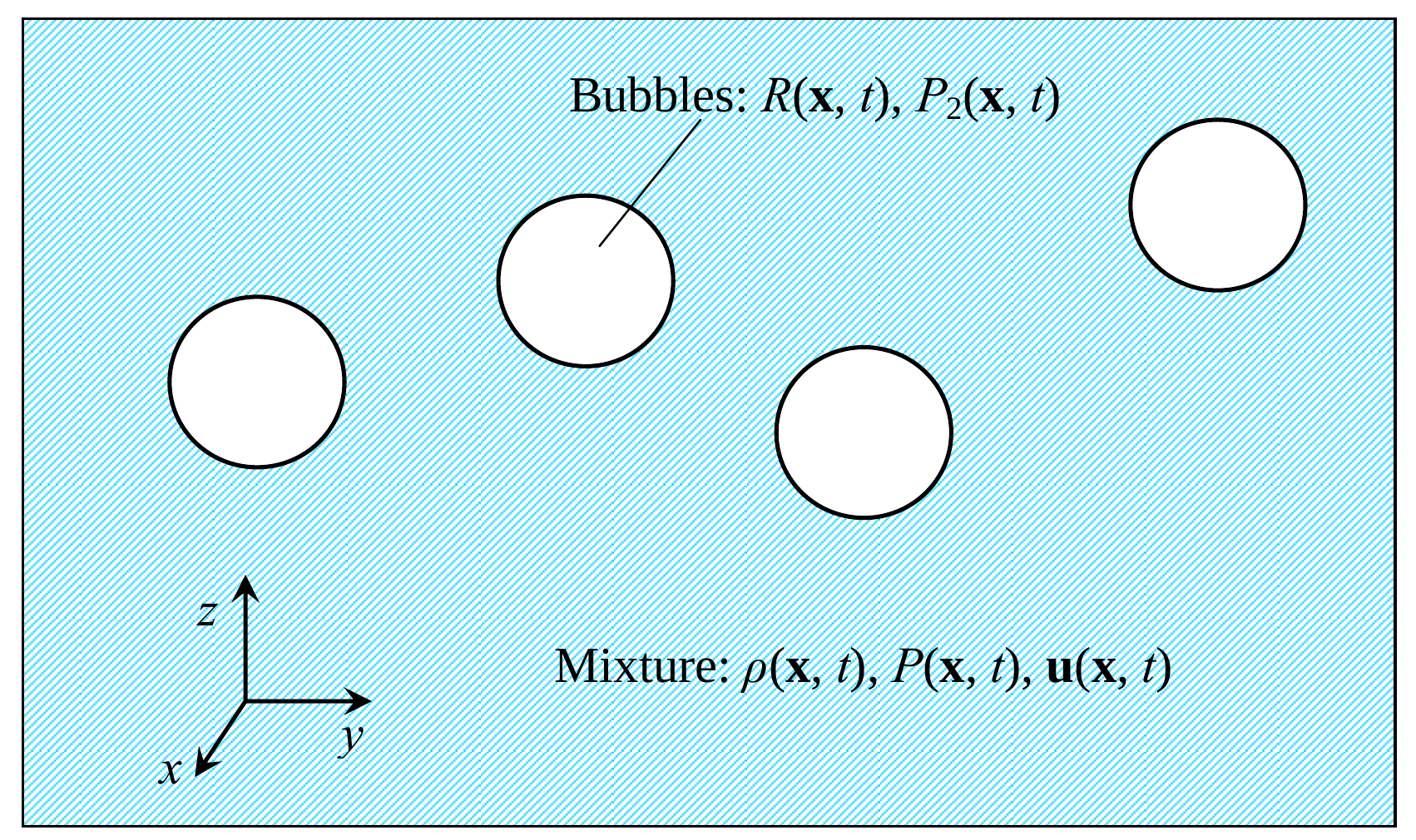}
\caption{\label{fig:bubb}The bubbly fluid.}
\end{figure}


\subsubsection{The three-dimensional PDE model}

So far, our model consists of two sets of equations. The first set of local equations is composed of ODEs: the Rayleigh-Plesset equation \eqref{eq:RPle}, the bubble thermodynamics equation \eqref{heatTransfer3}, and the algebraic mass balance equation \eqref{gasliquidCons} for the unknown local quantities
\begin{equation}\label{of:t}
 R(t) ,  P(t) ,  P_2(t) ,  \rho_2(t).
\end{equation}
The second set of equations is given by the the PDEs  \eqref{massconservation}, \eqref{Momcon} describing the dynamics of the density, velocity and pressure $\rho(\vec{x},t)$, $\vec{u}(\vec{x},t)$ and $P(\vec{x},t)$ of the mixture as a whole.

The two sets of equations can be put into a common framework if the local quantities \eqref{of:t} are assumed to change, for all bubbles, throughout the mixture flow domain, as functions of $\vec{x},t$. In order to preserve the Galilean invariance, which is essential in continuum mechanics, the time derivative ${d}/{dt}$ of any microscopic variable $\phi=\phi(\vec{x},t)$  \eqref{of:t} are naturally replaced by the \emph{total (material) derivative}
\beq \label{eq:matD3}
\matD_t\,\phi =\dfrac{\partial \phi}{\partial t} + \vec{u} \cdot \grad \,\phi,
\eeq
describing the rate of change of the respective quantity in the material frame of reference that moves with the flow.

Finally, the one-dimensional Galilei-invariant PDE model of a viscous bubbly fluid flow satisfying the above-described assumptions is given by
\begin{subequations} \label{GalInv3D}
\begin{equation}\label{GalInv3D:conti}
\pder{\rho}{t} + \div(\rho \vec{u}) = 0,
\end{equation}
\begin{equation}\label{GalInv3D:mom}
\rho \left(\pder{\vec{u}}{t}+ (\vec{u}\cdot\nabla)\vec{u}\right) + \grad\, P =\mu \Delta{\vec{u}} - \rho g\vec{k},
\end{equation}
\begin{equation} \label{GalInv3D:P}
P = P_2 - \rho_1\left(R\, \matD_t^2R + \dfrac{3}{2}\left(\matD_tR\right)^2 + \frac{4 \mu}{3 \rho_1}\dfrac{\matD_t R}{R}\right),
\end{equation}
\begin{equation} \label{GalInv3D:P2}
\matD_t P_2  + \dfrac{3\gamma P_2}{R}\matD_t R + \frac{3(\gamma-1){h}T_{0}}{R}{\left(\dfrac{P_2}{P_{0}}\frac{R^3}{{R_0}^3} - 1\right)} = 0 .
\end{equation}
\begin{equation} \label{GalInv3D:Rrho}
\left(1 - X + \dfrac{4}{3}{\pi}{N}R^3\rho_1\right)\rho - \rho_1 = 0.
\end{equation}
\end{subequations}
The system \eqref{GalInv3D} involves seven equations and seven (scalar) unknown physical fields described by the dependent variables $\rho$, $\vec{u}$, $P_2$, $P$, and $R$, which are functions of the independent variables $\vec{x}$ and $t$. In \eqref{GalInv3D},
\begin{equation} \label{GalInv:params}
\mu,~~\rho_1,~~\gamma,~~h,~~N,~~X,~~g  
\end{equation}
are constant material (constitutive) parameters that are determined by physical properties of the bubble-fluid mixture.

A \emph{well-posed problem} describing a specific model using the equations \eqref{GalInv} would involve application-specific boundary conditions on a finite, infinite, or periodic physical domain $V\subset \mathbb{R}^3$, and a set of initial conditions, given by, for example,
\begin{equation} \label{GalInv:ICs}
u\big|_{t = 0}=u_{0}(\vec{x}),\quad \rho \big|_{t = 0}=\rho_0(\vec{x}),\quad P\big|_{t = 0}= P_2\big|_{t = 0}=P_{0}(\vec{x}), \quad R\big|_{t = 0}=R_{0}(\vec{\vec{x}}).
\end{equation}

\subsubsection{One-dimensional reductions}

For models with symmetry, in the cases of flow parameters changing mostly in a single direction, it is natural to consider one-dimensional reductions of the full three-dimensional model \eqref{GalInv3D}. In this case, without loss of generality, the physical fields $\rho$, $u$, $P_2$, $P$, and $R$ are functions of the independent variables $x$ and $t$ only, and the PDE model becomes
\begin{subequations} \label{GalInv}
\begin{equation}
	\pder{\rho}{t} + \pder{}{x}(\rho u) = 0,
\end{equation}
\begin{equation}
	\pder{}{t}(\rho u) + \pder{}{x}(\rho u^2 + P) - \mu\pderr{u}{x} +g\rho= 0,
\end{equation}
\begin{equation} \label{GalInv:P}
P = P_2 - \rho_1\left(R\, \matD_t^2R + \dfrac{3}{2}\left(\matD_tR\right)^2 + \frac{4 \mu}{3 \rho_1}\dfrac{\matD_t R}{R}\right),
\end{equation}
\begin{equation} \label{GalInv:P2}
\matD_t P_2  + \dfrac{3\gamma P_2}{R}\matD_t R + \frac{3(\gamma-1){h}T_{0}}{R}{\left(\dfrac{P_2}{P_{0}}\frac{R^3}{{R_0}^3} - 1\right)} = 0 .
\end{equation}
\begin{equation} \label{GalInv:Rrho}
\left(1 - X + \dfrac{4}{3}{\pi}{N}R^3\rho_1\right)\rho - \rho_1 = 0,
\end{equation}
\end{subequations}
a model consisting of five equations and five unknown fields $\rho(x,t)$, $u(x,t)$, $P_2(x,t)$, $P(x,t)$, and $R(x,t)$.
For horizontal flows, one chooses $g = 0$, and for vertical flows with an upward-directed $x$ axis, $g>0$. The one-dimensional total derivative operator is given by
\beq \label{eq:matD}
\matD_t =\dfrac{\partial }{\partial t} + u\dfrac{\partial }{\partial x}.
\eeq

\begin{remark}\label{rem:galsol}
Since the system \eqref{GalInv} involves time only through the material time derivative $\matD_t$, all equations are Galilei-invariant. In particular, if
\[
\rho(x,t),~u(x,t),~ P(x,t),~ P_2(x,t),~R(x,t)
\]
solve the PDE system \eqref{GalInv}, then the quantities
\beq\label{eq:GalTr:newsol}
\barr
\rho^\dag(x,t)=\rho(x-ct,t),\\[1ex]
u^\dag(x,t)=u(x-ct,t)+c,\\[1ex]
P^\dag(x,t)=P(x-ct,t),\\[1ex]
P_2^\dag(x,t)=P_2(x-ct,t),\\[1ex]
R^\dag(x,t)=R(x-ct,t)
\earr
\eeq
provide a family of additional solutions holding for an arbitrary constant $c$. The solutions \eqref{eq:GalTr:newsol} describe the values of the physical fields observed in an inertial frame of reference moving in the negative $x$-direction at a  constant the speed $c$. [The same holds for the three-dimensional model \eqref{GalInv3D}, with the constant $c$ replaced by a vector $\vec{c}$.]
\end{remark}


\subsection{Dimensionless equations}

Classes of physical models involving constitutive parameters and/or functions may often be simplified using \emph{equivalence transformations}; in particular, constitutive parameters may be removed, and/or constitutive functions may be reduced to certain simpler forms (see, e.g., \cite{cheviakov2017symbolic} and references therein). In many cases, such simplifying transformations can be found by inspection; this is the case for the viscous bubble flow model above. We show how a class of scaling transformations (a general non-dimensionalization) leading to a dimensionless version of the model is used to substantially reduce the number of parameters. Alternatively, when scaling parameters are not ``chosen" but rather are prescribed by typical physical values, a ``physical" non-dimensionalization is obtained, involving well-known physical dimensionless parameters, as shown below.

We now derive two different non-dimensional versions of the multi-phase flow model at hand. For brevity, the results are given for the one-dimensional system \eqref{GalInv}; the results carry over to the full three-dimensional model \eqref{GalInv3D} in an obvious way.

\subsubsection{A general non-dimensionalization }\label{nondim:gen}

First, we wish to find a dimensionless version of the equations \eqref{GalInv} with a goal to remove as many constitutive parameters present in the system as possible. Consider a class of transformations
\beq\label{dimensionless}
\barr
	x= A_x\widetilde{x},  \qquad t = A_t\widetilde{t}, \qquad P = A_p{\widetilde{P}},  \qquad \rho = A_\rho\widetilde{\rho}, \\[2ex]
	u = A_u\widetilde{u}, \qquad  R =A_r\widetilde{R}, \qquad T = A_T\widetilde{T},
\earr
\eeq
where the constants $A_x$, $A_t$, $A_P$, $A_\rho$, $A_u$, $A_r$, $A_T$ retain the physical dimension of the respective variables, providing their ``characteristic" values based on dimensional considerations only, and the new variables $\widetilde{x}$, $\widetilde{t}$, etc. are dimensionless.

Upon the substitution of \eqref{dimensionless} into the PDE system \eqref{GalInv}, one requires
\[
\barr
A_x = A_t A_u, \qquad  A_t = \dfrac{\mu}{A_P}, \qquad A_P = \rho_1 \left(\dfrac{A_r}{A_t}\right)^2, \qquad  A_\rho &= \rho_1, \\[2.5ex]
A_u^2 = \dfrac{A_P}{A_\rho}, \qquad  A_r = \left({\dfrac{4}{3}\pi N \rho_1}\right)^{-{1}/{3}},\qquad A_T = \dfrac{A_r A_P}{h A_t}.
\earr
\]
Then the choice
\[
\barr
A_x = A_r =\left({\dfrac{4}{3}\pi N \rho_1}\right)^{-{1}/{3}}, \qquad A_t = \dfrac{\rho_1}{\mu} A_x^2, \qquad A_P =\dfrac{\mu^2}{\rho_1 A_x^2}, \\[2.5ex]
A_\rho = \rho_1, \qquad  A_u = \dfrac{\mu}{\rho_1 A_x}, \qquad A_T = \dfrac{4}{3}\dfrac{\pi\mu^3}{h \rho_1}
\earr
\]
leads to the elimination of several constant parameters \eqref{GalInv:params} of the model.
Specifically, the dimensionless version of the PDE system  \eqref{GalInv}, in terms of the unknowns $\widetilde{\rho}$, $\widetilde{u}$, $\widetilde{R}$, $\widetilde{P}$, $\widetilde{P}_2$ depending on $\widetilde{x}$ and $\widetilde{t}$, is given by
\begin{subequations} \label{ND}
 			\begin{equation} \label{massConsND}
				\pder{\widetilde{\rho}}{\widetilde{t}} + \pder{}{\widetilde{x}}(\widetilde{\rho} \widetilde{u}) = 0,
			\end{equation}
			\begin{equation} \label{MonConsND}
				\pder{}{\widetilde{t}}(\widetilde{\rho} \widetilde{u}) + \pder{}{\widetilde{x}}(\widetilde{\rho} \widetilde{u}^2 + \widetilde{P}) - \pderr{\widetilde{u}}{\widetilde{x}} +\kappa\widetilde{\rho}= 0,
			\end{equation} 	
			\begin{equation} \label{PfuncND}
				\widetilde{P} = \widetilde{P_2} - \left(\widetilde{R}\, \matD^2_{\widetilde{t}}\widetilde{R} + \frac{3}{2}(\matD_{\widetilde{t}}\widetilde{R})^2 + \frac{4}{3}\frac{\matD_{\widetilde{t}}\widetilde{R}}{\widetilde{R}}\right),
			\end{equation}
			\begin{equation} \label{F1funcaltND}
				\matD_{\widetilde{t}}\widetilde{P}_2 + 3\gamma \widetilde{P}_2\frac{\matD_{\widetilde{t}}\widetilde{R}}{\widetilde{R}} + 3(\gamma -1)\frac{\widetilde{T}_{0}}{\widetilde{R}}\left(\frac{\widetilde{P}_2}{\widetilde{P}_{0}}\frac{\widetilde{R}^3}{{\widetilde{R}_0}^3} - 1\right) = 0,
			\end{equation}
			\begin{equation} \label{F2funcND}
				 (1 - X + \widetilde{R}^3)\widetilde{\rho} - 1 = 0,
			\end{equation}
\end{subequations}
and involves only three dimensionless parameters: the gas mass fraction $X$, the adiabatic exponent $\gamma$, and the scaled gravity acceleration
\beq\label{eq:kappG}
\kappa = \frac{g \rho_1}{4 \pi N \mu^2}.
\eeq
For gravity-independent/horizontal flows, $\kappa =g = 0$.

\begin{remark} \label{SmallKappa}
For some materials, the scaled gravity acceleration term \eqref{eq:kappG} in the PDE \eqref{MonConsND} is rather small: $\kappa \ll 1$. In particular, this is the case the basilic magmas.
\end{remark}

The dimensionless version \eqref{ND} of the gas-fluid mixture flow model \eqref{GalInv} is useful for the general analysis of solution behaviour and the computation of exact and numerical solutions when no further approximations are required.

\subsubsection{A physical non-dimensionalization }\label{nondim:phys}

In Section \ref{nondim:gen}, the   dimensional values $A_i$ of the physical fields were computed from the requirement of elimination of as many constitutive parameters as possible.  The non-dimensionalizing rescaling can also be used for a different purpose, namely, to provide \emph{relative physical scales} of the terms in the equations, when the typical values of physical variables are \emph{known}. We now follow the second route, re-scaling the model equations \eqref{GalInv} according to the formulas
\beq\label{eq:nond:tr:phys}
\barr
	x = L x^*, \qquad t = \dfrac{L}{v_0} t^*, \qquad	P = A_P{P^*}, \qquad	P_2 = A_P{P_2^*},  \\[2ex]
	\rho = \rho_1 \rho^*, \qquad u = v_0 u^*, \qquad R =A_r R^*, \qquad 	T = A_T T^*,
\earr
\eeq
where $A_p$ is the characteristic pressure, $\rho_1$ is the density of the liquid, $L$ is the characteristic length, $v_0$ is the characteristic speed of the mixture, $R_0$ is the initial radius of the bubble, and $A_T$ is the characteristic temperature. In terms of dimensionless starred variables, one has
\[
\matD_{t^*} = \pder{}{t^*} + u^*(x^*,t^*)\pder{}{x^*}.
\]
The dimensionless version of the PDE system  \eqref{GalInv} arising from the transformation \eqref{eq:nond:tr:phys}, in terms of the dimensionless fields $\rho^*$, $u^*$, $R^*$, $P^*$, $P^*_2$ depending on ${x^*}$ and ${t^*}$, is given by
\begin{subequations}\label{physND}
\begin{equation} \label{Su_G_3}
	\pder{\rho^*}{t^*} + \pder{}{x^*}({\rho^*}u^*) = 0,
\end{equation}
\begin{equation} \label{Su_G_4}
	\pder{}{t^*}({\rho^*}u^*) + \pder{}{x^*}({\rho^*}{u^*}^2) + \Eu\, P^*_{x^*} - \frac{1}{\Re}\pderr{u^*}{(x^*)}  + \varkappa\rho^* = 0,
\end{equation}
\begin{equation} \label{physND:RP}
	P^*= P^*_2 -  \dfrac{\delta^2}{\Eu} \left(R^*\,\matD_{t*}^2R^* + \dfrac{3}{2}(\matD_{t^*}R^*)^2\right) - \dfrac{4}{3\,\Re\,\Eu}\dfrac{\matD_{t^*}R^*}{R^*},
\end{equation}
\begin{equation} \label{F1funcalt}
\matD_{t^*}P^*_2 + 3\gamma P^*_2\dfrac{\matD_{t^*}R^*}{R^*} + 3(\gamma-1)\dfrac{{W}}{\Eu\,\delta} \frac{T_{0}^*}{R^*}{\left(\frac{P_2^*}{P_{0}^*}\left(\dfrac{R^*}{R_0^*}\right)^3 - 1\right)}= 0,
\end{equation}
\begin{equation} \label{mixturecons}
	(1 - X + B {R^*}^3)\rho^* - 1= 0.
\end{equation}
\end{subequations}
The system \eqref{physND} involves several dimensionless constants: the Reynolds number
\[
\Re = \dfrac{\rho_1{v_0}{L}}{\mu}
\]
measuring the ratio of inertial forces to viscous forces, the
Euler  number
\[
\Eu = \dfrac{A_P}{\rho_1{v_0}^2}
\]
measuring the ratio of pressure forces to inertial forces, the bubble size to characteristic length ratio
\[
\delta = \dfrac{A_r}{L},
\]
the typical gas content in the mixture (gas mass per kilogram of mixture)
\[
\B = \dfrac{4}{3}\pi{N}{A_r}^3 \rho_1,
\]
the thermal constant 
\[
W = \frac{A_T h}{v_0\rho_1}, 
\]
and the gravity-related dimensionless constant
\[
\varkappa = \dfrac{g L}{ v_0^2},
\]
which vanishes for horizontal flows.

\begin{remark} For basaltic magmas, typical values for these constants at the magma chambers are as follows \cite{stolper1980melt}:
			\begin{align*}
				N &\sim 3\times{10^{4}}~\text{kg}^{-1}, & A_r &\sim 0.5\times10^{-3}~\text{m},\\
				\rho_1 &\sim 2800 ~{\text{kg}}/{\text{m}^3},& A_P &\sim 10^8~\text{Pa},\\
				h &\sim 1400~{\text{J}}/({\text{m}\cdot{\text{s}}\cdot{\text{K}}}),&A_T &\sim 1600~\text{K},\\
				v_0 &\sim 1~{\text{m}}/{\text{s}},& L &\sim 400~\text{m},\\
				\mu &\sim 100~\text{Pa}\cdot{\text{s}}.
			\end{align*}
For these parameter values, some terms in \eqref{physND} become much smaller than other ones. In particular the radius terms in \eqref{physND:RP} become six orders of magnitude smaller than the pressure terms.

\medskip We also list the typical values of the physical constants for magma flows in lava tubes\cite{JGRB:JGRB11445}:
			\begin{align*}
				N &\sim 6~\text{kg}^{-1},& A_r &\sim 0.01 ~\text{m},\\
				\rho_1 &\sim 2800 ~{\text{kg}}/{\text{m}^3},& A_P &\sim 10^5~\text{Pa},\\
				h &\sim 1400~{\text{J}}/({\text{m}\cdot{\text{s}}\cdot{\text{K}}}),&A_T &\sim 1200~\text{K},\\
				v_0 &\sim 1~{\text{m}}/{\text{s}},& L &\sim 0.1~\text{m},\\
				\mu &\sim 100~\text{Pa}\cdot{\text{s}}.
			\end{align*}

\medskip As an industrial-related application of the multiphase model of interest, one can consider machine oil with bubbles; some typical values of the physical constants are
			\begin{align*}
				N &\sim 3000~\text{kg}^{-1},& A_r &\sim 0.001 ~\text{m},\\
				\rho_1 &\sim 900 ~{\text{kg}}/{\text{m}^3},& A_P &\sim 10^6~\text{Pa},\\
				h &\sim 1400~{\text{J}}/({\text{m}\cdot{\text{s}}\cdot{\text{K}}}),& A_T &\sim 515~\text{K},\\
				v_0 &\sim 1~{\text{m}}/{\text{s}},& L &\sim 0.1 \text{m},\\
				\mu &\sim 0.3~\text{Pa}\cdot{\text{s}}.
			\end{align*}
The dimensionless parameters corresponding to the above cases are summarized in Table \ref{tab:1}. It is evident that in the corresponding dimensionless equations \eqref{physND}, coefficients at different terms may have vastly varying magnitudes.
\begin{table}[h]
\centering
\begin{tabular}{|l|l|l|l|l|l|}
\hline
Case            & Eu      & Re      & $\delta$          & $B$    & $W$   \\ \hline
Magma Chamber   & $36000$ & $11000$ & $1\times 10^{-6}$ & $0.04$ & $800$ \\ \hline
Lava Flow Field & $36$    & $3$     & $0.1$             & $0.07$ & $600$ \\ \hline
Machine Oil     & $111$   & $300$   & $0.01$            & $0.01$ & $800$ \\ \hline
\end{tabular}
\caption{Typical dimensionless parameters for different applications.}
\label{tab:1}
\end{table}
\end{remark}

\subsection{Equilibrium and traveling wave solutions} \label{equilibriumSection}

We now seek equilibrium solutions
\begin{equation}\label{afs:equil:dimless:1}
\rho=\rho_e(x),\quad u= u_e(x) , \quad R=R_e(x),\quad P=P_e(x), \quad P_2=P_{2e}(x)
\end{equation}
of the dimensionless bubbly fluid equations \eqref{ND}, with tildes omitted for brevity. The substitution of \eqref{afs:equil:dimless:1} into \eqref{ND} and a brief computation lead to the following result.
\begin{proposition}\label{prop:eqsol:1}
The model \eqref{ND} admits equilibrium solutions \eqref{afs:equil:dimless:1} of the following form:
\begin{equation}\label{afs:equil:dimless:u}
\barr
\rho_e(x) = \dfrac{K_e}{u_e(x)}, \\[2ex]
P_e(x) = P_0 +\der{u_e(x)}{x} - K_e u_e(x) + \displaystyle \kappa\int\rho_e(x) \,dx, \\[2ex]
P_{2e}(x) = P_0\left(K_e (X-1) + u_e(x)\right)^{-\gamma} ,\\[2ex] R^3_e(x) = \dfrac{u_e(x)}{K_e} + X - 1,
\earr
\end{equation}
where $u_e(x)$ is an arbitrary solution of the above equations and the following ODE:
\begin{equation}\label{prop:eqsol:1:sol}
\barr
P_e(x) = P_{2e}(x) - u_e(x)\left(R_e(x)\der{^2 R_e(x)}{x^2} + \der{u_e(x)}{x} R_e(x)\der{R_e(x)}{x} \right.\\[2ex]
 \qquad \qquad +\left.\dfrac{3}{2}\left(\der{R_e(x)}{x}\right)^2 + \dfrac{4}{3 R_e(x)} \der{R_e(x)}{x}\right).
\earr
\end{equation}

\end{proposition}

\medskip We separately consider the important case $u_e = 0 $, the following equilibrium solution arises.
\begin{proposition}\label{prop:eqsol:2}
The model \eqref{ND} admits equilibrium solutions of the form \eqref{afs:equil:dimless:1} for the equilibrium quantities given by
\begin{equation}\label{afs:equil:dimless:u_0}
\barr
u_e(x) = 0,\quad
P_e(x) = P_0 + \displaystyle\kappa \int \rho_e(x) d x,   \quad
P_{2e}(x) = P_e(x), \quad
R^3_e(x)=\dfrac{1}{\rho_e(x)}+X-1,
\earr
\end{equation}
where $\rho_e(x)$ is an arbitrary equilibrium  density distribution.
\end{proposition}
\begin{remark}
From \eqref{afs:equil:dimless:u_0} one observes that for vertical flows $(\kappa > 0)$ described by the model, as the bubbles rise, the bubble radius does not increase. This is a consequence of the model assumption of no mass exchange between the two phases.
\end{remark}


\subsection{Traveling wave solutions}\label{sec:trv}
Since the PDE system \eqref{ND} is invariant with respect to space and time translations, it admits traveling wave solutions of the form
\beq\label{eq:trw:ansatz}
f(x,t) = f(\xi),\qquad  \xi = x - c t,\qquad c=\const,
\eeq
where $f(x,t)$ stands for each of the physical fields $\rho, u, P, P_2, R$. The substitution of the traveling wave ansatz \eqref{eq:trw:ansatz} converts the PDEs in \eqref{ND} into ODEs with the independent variable $\xi$; the ODE system may subsequently be solved to find the traveling wave solutions. However, since the model \eqref{ND} is Galilei-invariant, the ODEs of the traveling wave ansatz will essentially coincide with the equilibrium solutions modified by a Galilei transformation \eqref{eq:GalTr:newsol} (see Remark \ref{rem:galsol}). The following result holds.

\begin{proposition} \label{cololl:galil}
For any equilibrium configuration $\rho_e(x), u_e(x), P_e(x), P_{2e}(x), R_e(x)$ solving \eqref{afs:equil:dimless:u}, \eqref{prop:eqsol:1:sol} or \eqref{afs:equil:dimless:u_0}, the model \eqref{ND} admits a family of time-dependent exact solutions given by
\begin{equation}\label{eq:galil:sols}
\rho_e(x-ct),~u_e(x-ct)+c,~P_e(x-ct),~P_{2e}(x-ct),~R_e(x-ct)
\end{equation}
for an arbitrary $c=\const$.
\end{proposition}

Proposition \ref{cololl:galil} yields explicit traveling wave solutions for any given equilibrium solution.

\section{Perturbation analysis} \label{pertAnalysis}


As it is common for nonlinear models, it is not feasible to derive an exact closed-form general solution of the full bubbly fluid PDE model \eqref{ND}. We now seek its approximate solutions using an extended version of the Su-Gardner procedure \cite{su1969korteweg}. This procedure applies to Galilean-invariant models which contain the continuity and momentum balance equations ($g =0$), as well as possibly other algebraic or differential equations, involving the corresponding number of additional dependent variables $f$:
\begin{subequations}
    \begin{equation}
    \pder{\rho}{t} + \pder{}{x}(\rho u ) = 0,
    \end{equation}
    \begin{equation}
    \pder{}{t}(\rho u) + \pder{}{x}(\rho u^2 + P) = 0,
    \end{equation}
    \begin{equation}\label{eq:SG:Peq}
      P = P(f,\rho,u,f_x,\rho_x,u_x,f_t,\ldots),
    \end{equation}
    \begin{equation}
      F(f,\rho,u,f_x,\rho_x,u_x,f_t,\ldots) = 0,
    \end{equation}
\end{subequations}
In our case (i.e., the model \eqref{ND} with tildes omitted), $f$ includes both of the gas state variables $R$ and $P_2$, the equation \eqref{eq:SG:Peq} is given by \eqref{PfuncND}, and $F$ has two components given by \eqref{F1funcaltND} and
\eqref{F2funcND}. The Su-Gardner approach \cite{su1969korteweg} maintains that for a Galilei-invariant system with a ``weak" nonlinearity, one can take the long-wavelength approximation
\begin{equation}
\xi = \epsilon^{\alpha}\left(x - a_0 t\right), \qquad \tau = \epsilon^{\alpha + 1}t,
\end{equation}
and expand the state variables asymptotically about the constant equilibrium state:
\begin{equation}\label{eq:SG:series}
 \begin{split}
   \rho &= \rho_0 + \epsilon \rho^{(1)} + \epsilon^2 \rho^{(2)}+ \ldots ,\\
   f &= f_0 + \epsilon f^{(1)} + \epsilon^2 f^{(2)}+ \ldots ,\\
   u &= 0 + \epsilon u^{(1)} + \epsilon^2 u^{(2)}+ \ldots .
 \end{split}
\end{equation}
Here $\epsilon\ll 1$ is an arbitrary small parameter controlling the magnitude of the perturbation, $a_0=\const$ is the wave speed, $\alpha=\const>0$ is a power parameter, and $\tau$ is the ``slow time" variable.

After the substitution of the ansatz \eqref{eq:SG:series} into the governing equations, every power of $\epsilon$ up to a specified precision is required to vanish independently.


\subsection{An asymptotic expansion with  generalized power series }

Consider again the dimensionless bubbly fluid model \eqref{ND}, with tildes again omitted everywhere for brevity, in the case of horizontal flows and other flows where the gravity is negligible: $(\kappa = 0)$. Suppose that the state variables $\rho(x,t)$, $u(x,t)$, $P(x,t)$, $P_2(x,t))$, $R(x,t)$ can be represented as a generalized power series in terms of some small parameter $\epsilon$:
\begin{equation} \label{expansion}
\begin{split}
\rho &= \rho_0 + \epsilon^q \rho^{(1)} + \epsilon^{q+k} \rho^{(2)} + \epsilon^{q+2 k} \rho^{(3)} +\ldots,\\
u &=  0 + \epsilon^q u^{(1)} + \epsilon^{q+k} u^{(2)} +\epsilon^{q+2 k} u^{(3)} + \ldots,\\
R &= R_0 + \epsilon^q R^{(1)} + \epsilon^{q+k} R^{(2)} + \epsilon^{q+2 k} R^{(3)} +\ldots,\\
P &= P_0 + \epsilon^q P^{(1)} + \epsilon^{q+k} P^{(2)} + \epsilon^{q+2 k} P^{(3)} +\ldots,\\
P_2 &= P_0 + \epsilon^q P_2^{(1)} + \epsilon^{q+k} P_2^{(2)} + \epsilon^{q+2 k} P_2^{(3)} +\ldots\; .
\end{split}
\end{equation}
To account for the slow variation of the wave-form, we introduce Gardner-Morikawa coordinate transformation  \cite{kudryashov2006nonlinear} of the independent variables:
\begin{equation} \label{longWave}
\xi = \epsilon^{\alpha}\left(x - a_0 t\right), \qquad \tau = \epsilon^{\alpha + \beta}t.
\end{equation}
In the above formulas, $a_0$ is an arbitrary speed at which the reference frame is moving, $\alpha$, $\beta$, $q$ and $k$ are constant positive power parameters, and $\tau$ is the ``slow" time. In \eqref{expansion}, it is assumed that $\rho_0$, $R_0$, and $P_0$ are components of a constant equilibrium solution, and the perturbation series coefficients are functions of $(\xi,\tau)$.

\begin{remark}
We note that truncated series \eqref{expansion} represent approximate solutions of the full model \eqref{ND} to the desired order of $\epsilon$. The approximation based on \eqref{expansion} and \eqref{longWave} leads to new solutions to \eqref{ND} rather than an approximation of equilibrium solutions \eqref{afs:equil:dimless:u_0} translated through a Galilean transformation. Indeed, in the case of the equilibrium solution \eqref{afs:equil:dimless:u}, $u_e=u_e(x)$. For the equilibrium solution \eqref{afs:equil:dimless:u_0},  $u_e=0$; after a Galilei transformation, one has $u(x,t)=c\ne 0$. At the same time, in the Su-Gardener approximation \eqref{expansion}, $u(x,t) \sim  0 + \epsilon^q u^{(1)}(x,t)$, with $u^{(1)}(x,t)\ne 0$.
\end{remark}

The perturbation expansions \eqref{expansion} about the equilibrium solution \eqref{afs:equil:dimless:u_0} are required to annihilate the lowest-order terms of each of the equations \eqref{ND}. We note that $\epsilon^{\alpha + q}$ is a common factor in the continuity and momentum conservation equations, and $\epsilon^{q}$ factors out from the three remaining DEs. Setting to zero the lowest-order terms, we obtain the following relationships:
\begin{subequations} \label{lowestOrd}
\begin{equation}
\pder{}{\xi}u^{(1)} = \frac{a_0}{\rho_0}\pder{}{\xi}\rho^{(1)},
\end{equation}
\begin{equation}
P^{(1)} = P_2^{(1)},
\end{equation}
\begin{equation}
P_2^{(1)} = -\frac{3P_0}{R_0}R^{(1)},
\end{equation}
\begin{equation}
R^{(1)} = -\frac{(R_0^3-X+1)^2}{3 R_0} \rho^{(1)},
\end{equation}
\begin{equation}
\pder{}{\xi}P^{(1)} = \rho_0 a_0 \pder{}{\xi}u^{(1)}.
\end{equation}
\end{subequations}
The above equations involve a constraint on $a_0$; it can be shown that
\begin{equation} \label{a_0}
a_0 = {R_0^{-3/2} P_0^{1/2}}(R_0^3-X+1).
\end{equation}
This choice of $a_0$ satisfies \eqref{lowestOrd} for an arbitrary set of positive constants $\alpha$, $\beta$, $q$ and $k$.

We now consider several particular cases in which the bubbly fluid model \eqref{ND} is asymptotically equivalent to a \emph{single nontrivial PDE}, in the sense that $O(\epsilon^q)$ terms of the density expansion $\rho^{(1)}(\xi,\tau)$ in \eqref{expansion} will be governed by a PDE, and $O(\epsilon^q)$ terms of other fields are found from \eqref{lowestOrd}.
[We note that it was possible to identify three such cases; in principle, more cases can exist for other relations between Gardner-Morikawa exponents $\alpha$, $\beta$, $q$, and $k$.]

For each such case, one immediately obtains \emph{an approximate solution} of the bubble fluid model \eqref{GalInv} (through its dimensionless version \eqref{ND}), given by the lowest two terms of the expansion \eqref{expansion}.

\subsection{The Burgers equation case}\label{sec:ND:Burg}

First, consider the case $\alpha = \beta = q = k$. For this case, the second-lowest-order terms of the DEs \eqref{ND} yield the following relations between second-order and first-order perturbations \eqref{expansion}:
\begin{subequations}\label{eq:Burgpara:O2}
\begin{equation}
\pder{}{\xi}u^{(2)}  =  \frac{a_0}{\rho_0}\pder{}{\xi}\rho^{(2)}- \frac{a_0}{\rho_0}\left(\pder{}{\tau}\rho^{(1)} + \frac{2}{\rho_0}\rho^{(1)}\pder{}{\xi}\rho^{(1)}\right),
\end{equation}
\begin{equation}
P^{(2)}= P_2^{(2)} + \frac{4 a_0}{3 R_0} \pder{}{\xi}R^{(1)},
\end{equation}
\begin{equation}
P_2^{(2)} = -\frac{3P_0}{R_0}R^{(2)} + \frac{a_0 P_0^2}{T_0}\pder{}{\xi}R^{(1)} + \frac{6 P_0}{R_0^2}\left(R^{(1)}\right)^2,
\end{equation}
\begin{equation}
R^{(2)} = -\frac{(R_0^3-X+1)^2}{3 R_0} \rho^{(2)} + \frac{(2 R_0^3+X-1)(R_0^3-X+1)^3}{9 R_0^5} \left(\rho^{(1)}\right)^2.
\end{equation}
\end{subequations}
Solving the above system amended with \eqref{MonConsND} for $\rho^{(1)}$, we find that the first-order density perturbation $\rho^{(1)}(x,t)$ satisfies the well-known \emph{Burgers equation}:
\begin{equation} \label{BurgersEq}
\pder{\rho^{(1)}}{\tau} = A \pderr{\rho^{(1)}}{\xi} + B\rho^{(1)}\pder{\rho^{(1)}}{\xi},
\end{equation}
where the constants $A$, and $B$ are given by
\begin{equation} \label{BurgConst}
\begin{split}
A &= {\frac { \left[ {P_0}^{2}{R_0}^{4}+\tfrac{13}{3}{R_0}^{3
}T_0+\left({P_0}^{2} R_0 + \tfrac{4}{3}\,T_0\right) (1-X)  \right]  \left({R_0}^{3}-X+1 \right) }{6{R_0}^{3}T_0}}, \\
B &= -{\frac {{P_0}^{1/2} \left( {R_0}^{3}-X +1\right) ^{3}}{{R_0}^{9/2}}}.
\end{split}
\end{equation}

We have obtained the following result.
\begin{proposition}
In the case of Gardner-Morikawa exponents $\alpha = \beta = q = k$, the lowest-order $\epsilon$-dependent terms of the equations \eqref{ND} are satisfied by the first two terms of the expansions \eqref{expansion}, where $\rho^{(1)}$ is an arbitrary solution of the Burgers equation \eqref{BurgersEq}, and the remaining state variable perturbations $u^{(1)}$, $P^{(1)}$, $P_2^{(1)}$, $R^{(1)}$ , $u^{(2)}$, $P^{(2)}$, $P_2^{(2)}$ and $R^{(2)}$ are determined by the relationships \eqref{lowestOrd} and \eqref{eq:Burgpara:O2}.
\end{proposition}

\begin{remark}
In a similar manner, one can analyze relationships between higher-order perturbations $\rho^{(2)}$, $u^{(2)}$, $P^{(2)}$, $P_2^{(2)}$ and $R^{(2)}$, etc. In particular, one can show that for the current choice of Gardner-Morikawa exponents, $\rho^{(2)}$ satisfies a modified diffusion equation
\[
C_1\pder{\rho^{(2)}}{\tau} = C_2\pderr{\rho^{(2)}}{\xi} + F_1\pder{\rho^{(2)}}{\xi} +F_2\rho^{(2)} + F_3,
\]
where $C_1, C_2=\const$, and the coefficients $F_1$, $F_2$, $F_3$ depend on the previous-order perturbation $\rho^{(1)}$.
\end{remark}

\medskip\noindent{\textbf{Example.}} As an illustration, we take $\alpha = \beta = q = k=1$, and employ a famous exact explicit kink-type traveling wave solution of the Burgers equation
\begin{equation} \label{BurgSol}
\rho^{(1)}(\xi,\tau)= \frac{2 c A}{B} \tanh(C_2\tau + c \xi - C_1)+\frac{C_2}{B c},
\end{equation}
$C_1, C_2, c=\const$, to construct approximate solutions of the bubbly flow equations \eqref{ND}. Indeed, the formula \eqref{BurgSol} together with the first-order perturbations $u^{(1)}$, $P^{(1)}$, $P_2^{(1)}$ and $R^{(1)}$ given by  \eqref{lowestOrd}  yield the first two terms of the solution expansion \eqref{expansion} of the PDE model \eqref{ND}. [It is straightforward to verify that these approximate solutions satisfy the equations \eqref{ND} up to an $\mathcal{O}(\epsilon^2)$ error.]

\begin{figure}[H]
        \includegraphics[width=0.5\textwidth]{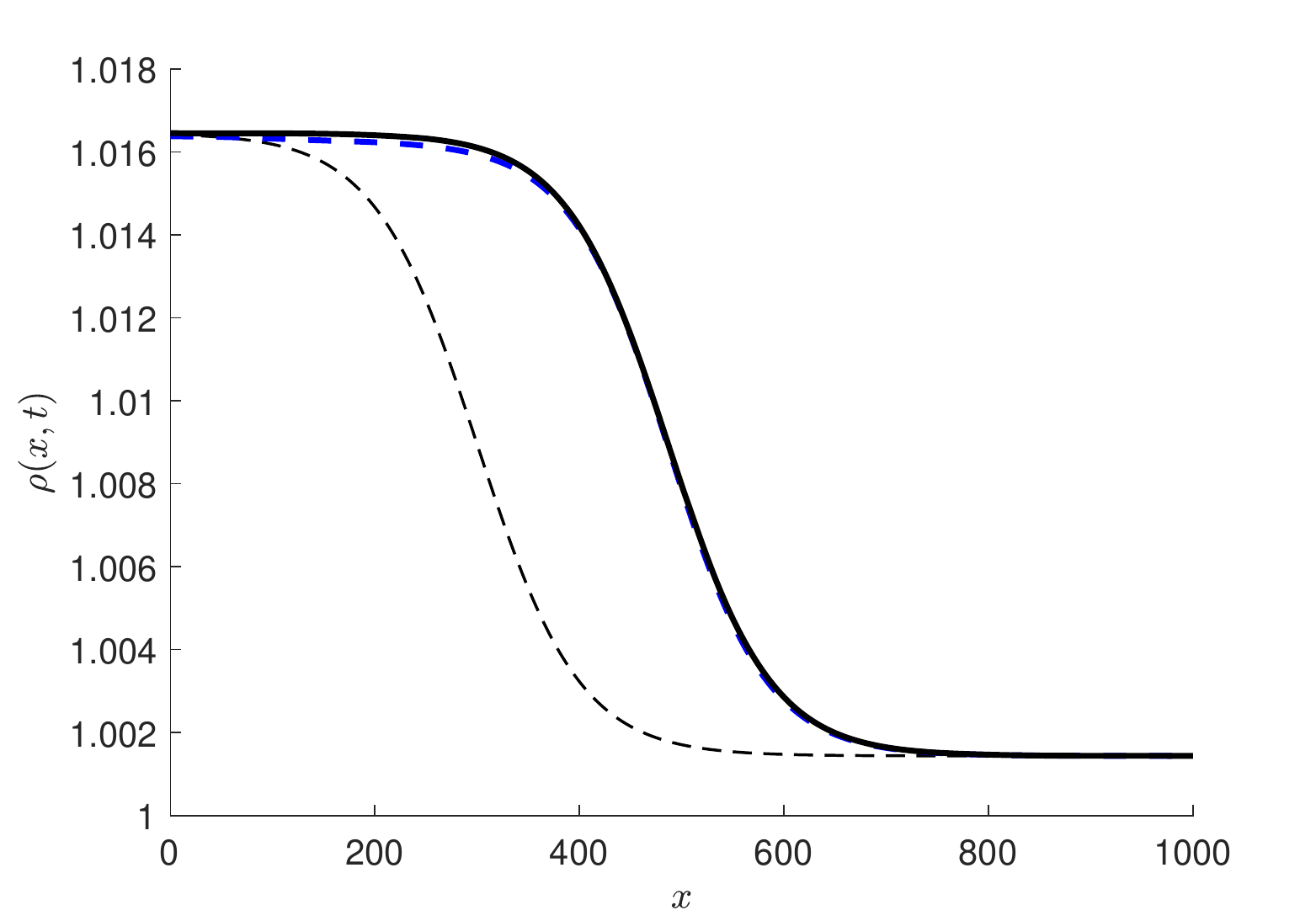}
        \includegraphics[width=0.5\textwidth]{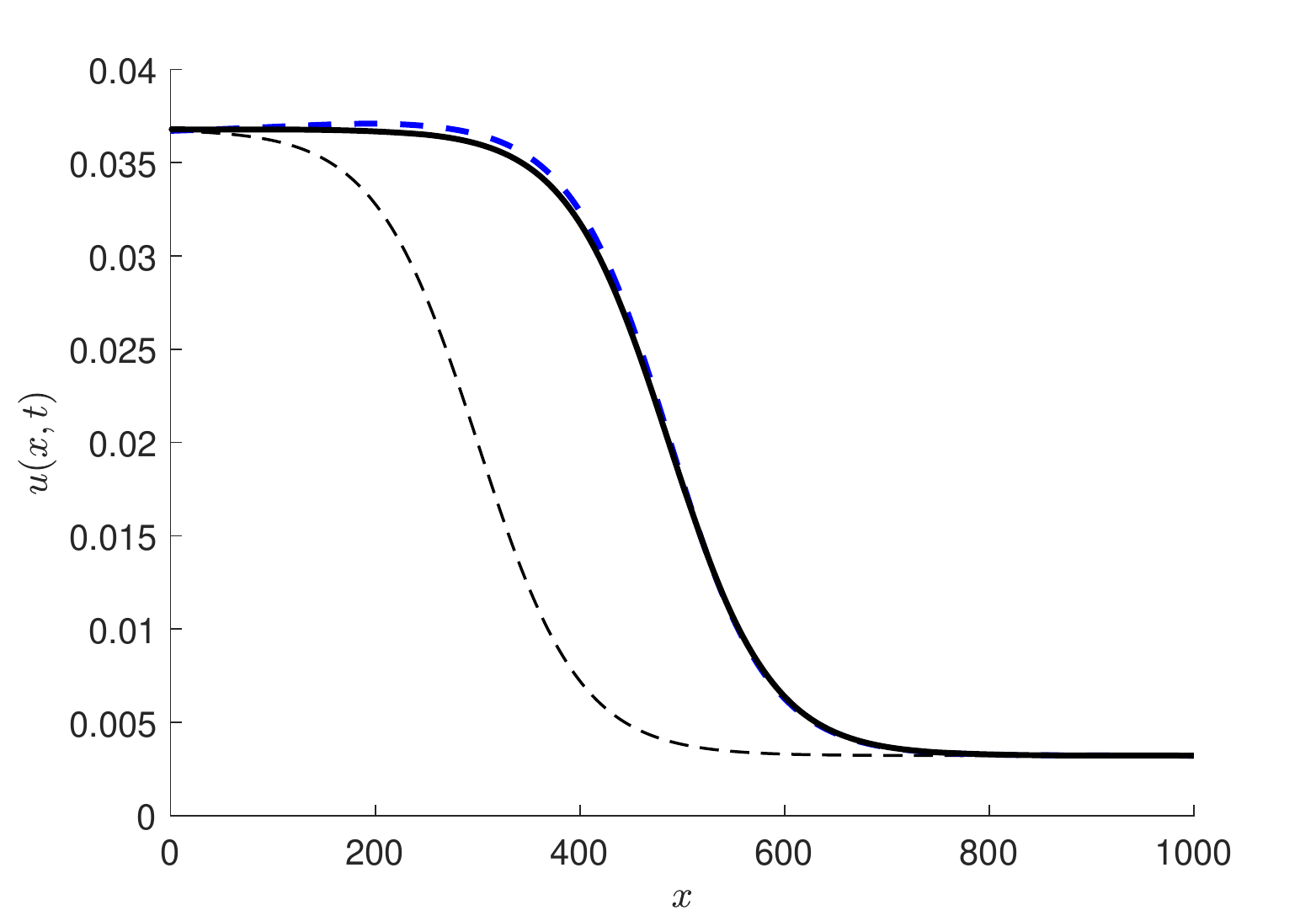}
        \includegraphics[width=0.5\textwidth]{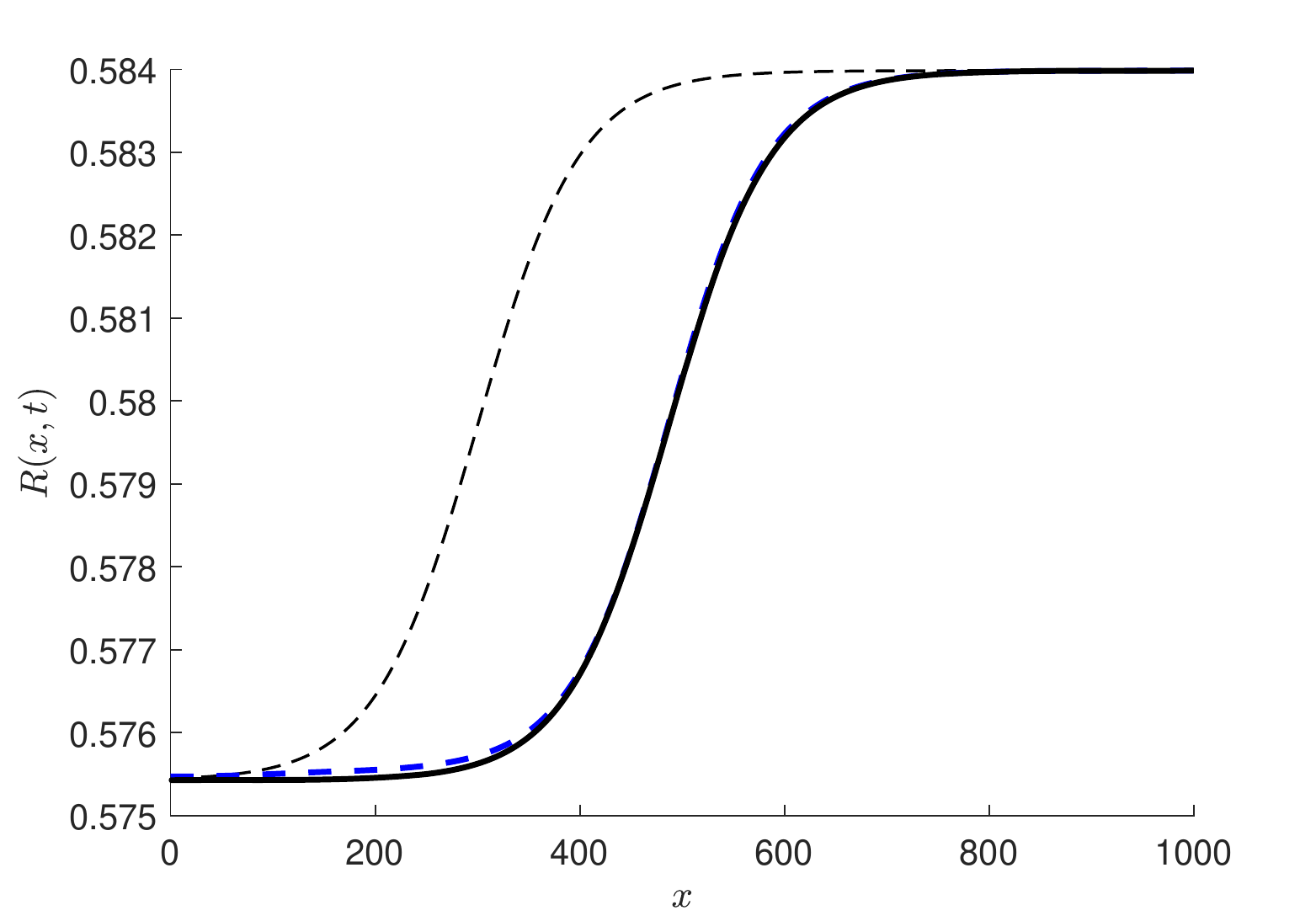}
\caption{\label{fig:Burg} A dimensionless approximate solution components $\rho(x,t), u(x,t), R(x,t)$ of the model equations \eqref{ND} given by the sum of $\mathcal{O}(1)$ and $\mathcal{O}(\epsilon)$ terms of the expansion \eqref{expansion}, for the parameters \eqref{eq:param:sol1} (tildes omitted), compared to a numerical solution of the full bubbly fluid model \eqref{ND}. The dashed line denotes the $tanh$-shaped initial condition. The thick black (solid) line denotes the approximate solution curves at the dimensionless time $t=80$. The thick blue (dashed) line denotes the numerical solution curves at $t=80$.}
\end{figure}

As a specific example, we compute two lowest-order terms of the terms of the approximate solution expansion \eqref{expansion} for the parameter values
\beq\label{eq:param:sol1}
\epsilon =0.01, \quad c= 1, \quad C_1 = 300, \quad C_2 = -20,  \quad \rho_e = 1, \quad P_0 = 1, \quad T_0 = 1.
\eeq
For these values, one obtains, in particular,
\[
a_0\simeq 2.2361,\quad A\simeq 4.1969, \quad B\simeq -11.1803,  \quad R_0\simeq 0.5848.
\]
The graphs of thus approximated physical fields $\rho(x,t), u(x,t), R(x,t)$ are given in Figure \ref{fig:Burg}. [We remind that throughout this section, including Figure \ref{fig:Burg}, the dimensionless fields and variables are considered, given by \eqref{dimensionless}. All tildes are omitted. For instance, by $\rho(x,t)$ we mean $\widetilde{\rho}(\widetilde{x},\widetilde{t})$, etc.]

In order to verify the quality of the approximation provided by the above-described truncated series approximate solution, we use a numerical solution of the full bubbly fluid model \eqref{ND} (see Appendix \ref{Sec:Num}), with the same initial condition, and compare the two at the dimensionless time $t=80$. A close agreement is observed.

 \subsection{The diffusion equation case}

Another interesting case is given by the choice $\alpha = \beta =  {q}/{2}=k$. Setting to zero the second-lowest-order terms in the PDEs yields
\begin{subequations}\label{eq:Heatpara:O2}
\begin{equation}
\pder{}{\xi}u^{(2)}  =  \frac{a_0}{\rho_0}\pder{}{\xi}\rho^{(2)} - \frac{1}{\rho_0}\pder{}{\tau}\rho^{(1)},
\end{equation}
\begin{equation}
P^{(2)}= P_2^{(2)} + \frac{4 a_0}{3 R_0} \pder{}{\xi}R^{(1)},
\end{equation}
\begin{equation}
P_2^{(2)} = -\frac{3P_0}{R_0}R^{(2)} + \frac{a_0 P_0^2}{T_0}\pder{}{\xi}R^{(1)},
\end{equation}
\begin{equation}
R^{(2)} = -\frac{(R_0^3-X+1)^2}{3 R_0} \rho_{(2)}.
\end{equation}
\end{subequations}
Using \eqref{MonConsND}, we find that the lowest-order density perturbation $\rho^{(1)}$ is governed by the linear diffusion equation
\begin{equation}\label{eq:rho:heat}
\pder{}{\tau}\rho^{(1)} = A\pderr{}{\xi}\rho^{(1)},
\end{equation}
where the constant coefficient $A$ is given by
\begin{equation}
A = \frac{\left[(P_0^2R_0 +\tfrac{4}{3}T_0) (1-X) +P_0^2R_0^4+\tfrac{13}{3}R_0^3 T_0 \right](R_0^3-X+1)}{6  R_0^3 T_0 }.
\end{equation}
We have obtained the following result.
\begin{proposition}
In the case of Gardner-Morikawa exponents $\alpha = \beta = k = {q}/{2}$, the lowest-order $\epsilon$-dependent terms of the equations \eqref{ND} are satisfied by the first two terms of the parameter expansions \eqref{expansion}, where $\rho^{(1)}$ is an arbitrary solution of the linear diffusion equation \eqref{eq:rho:heat}, and the remaining state variable perturbations $u^{(1)}$, $P^{(1)}$, $P_2^{(1)}$, $R^{(1)}$ , $u^{(2)}$, $P^{(2)}$, $P_2^{(2)}$ and $R^{(2)}$ are determined by the relationships \eqref{lowestOrd} and \eqref{eq:Heatpara:O2}.
\end{proposition}

\subsection{The Korteweg-de Vries equation case}\label{sec:pert:KdV}

Another exponent relationship leading to a nontrivial single PDE governing the leading-order perturbations is given by $\alpha = k = {\beta}/{2} = {q}/{2}$ (cf. \cite{Kud_Sine}). In this case, setting to zero the lowest-order terms $\epsilon$ yields the relations
\begin{subequations}\label{eq:KdVpara:O2}
\begin{equation}
\pder{}{\xi}u^{(2)}  =  \frac{a_0}{\rho_0}\pder{}{\xi}\rho^{(2)},
\end{equation}
\begin{equation}
P^{(2)}= P_2^{(2)} + \frac{4 a_0}{3 R_0} \pder{}{\xi}R^{(1)},
\end{equation}
\begin{equation}
P_2^{(2)} = -\frac{3P_0}{R_0}R^{(2)} + \frac{a_0 P_0^2}{T_0}\pder{}{\xi}R^{(1)},
\end{equation}
\begin{equation}
R^{(2)} = -\frac{(R_0^3-X+1)^2}{3 R_0} \rho_{(2)}.
\end{equation}
\end{subequations}
Repeating the steps from the other cases, we arrive at a constraint
\[
(R_0^3-X+1)^2\frac{\left((X-1)P_0^2R_0 -P_0^2R_0^4-\frac{13}{3}R_0^3T_0 + \frac{4}{3}T_0(X-1)\right)P_0^{1/2}}{R_0^{9/2}T_0} \pderr{}{\xi}\rho^{(1)}= 0.
\]
The above equation is satisfied by $\rho^{(1)} = A\xi + B$, which fails to be small, or by a specific choice of $T_0$:
\begin{equation}\label{kdvconstraint}
T_0 = -\frac{3P_0^2R_0(R_0^3-X+1)}{13R_0^3-4X+4}.
\end{equation}
Now examining the second-lowest order of $\epsilon$, one has
\begin{subequations}
\begin{equation}
\pder{}{\xi}u^{(3)}  =  \frac{a_0}{\rho_0}\pder{}{\xi}\rho^{(3)} -2 \dfrac{a_0}{\rho_0^2} \rho^{(1)}\pder{}{\xi}\rho^{(1)} - \frac{1}{\rho_0}\pder{}{\tau}\rho^{(1)},
\end{equation}
\begin{equation}
P^{(3)} = P_2^{(3)} + \frac{4 a_0}{3 R_0} \pder{}{\xi}R^{(2)} - R_0 a_0^2 \pderr{}{\xi}R^{(1)},
\end{equation}
\begin{equation}
R^{(3)} = -\frac{(R_0^3-X+1)^2}{3 R_0} \rho^{(3)} + \frac{(2 R_0^3+X-1)(R_0^3-X+1)^3}{9 R_0^5} \left(\rho^{(1)}\right)^2.
\end{equation}
\end{subequations}
Simplifying the above equations, we determine that $\rho^{(1)}$ satisfies the \emph{Korteweg-de Vries (KdV) equation}
\begin{equation}\label{eq:KdV1}
A\pder{\rho^{(1)}}{\tau} = B \pder{^3\rho^{(1)}}{\xi^3}+ C\rho^{(1)}\pder{\rho^{(1)}}{\xi},
\end{equation}
where $A$, $B$, and $C$ are constant parameters given by
\begin{equation}
\begin{split}
A &= {P_0}^{1/2}R_0^5\,(\gamma-1),\\
B &= \frac{1}{6}{R_0}^{1/2}\left(R_0^3-X+1\right)\left(P_0 R_0^2\left(R_0^3-X+1\right)^2(\gamma-1)-\frac{16}{27}\left(\frac{13}{4} R_0^3-X+1\right)^2\right) , \\
C &=  P_0{R_0}^{1/2}\,\left(R_0^3-X+1\right)^3(\gamma-1) .
\end{split}
\end{equation}
The following result holds.
\begin{proposition}
For the case of Gardner-Morikawa exponents $\alpha = k = {\beta}/{2} = {q}/{2}$, the lowest-order $\epsilon$-dependent terms of the equations  \eqref{ND} are satisfied by the first two terms of the parameter expansions \eqref{expansion}, where $\rho^{(1)}$ is an arbitrary solution of the KdV equation \eqref{eq:KdV1}, and the remaining state variable perturbations $u^{(1)}$, $P^{(1)}$, $P_2^{(1)}$, $R^{(1)}$ , $u^{(2)}$, $P^{(2)}$, $P_2^{(2)}$ and $R^{(2)}$ are determined by the relationships \eqref{lowestOrd} and \eqref{eq:KdVpara:O2}.
%
%
\end{proposition}
\begin{remark}
Since physically, $P_0, R_0>0$ and $0<X <1$, from \eqref{kdvconstraint} it follows that in this case, the initial temperature must be negative, $T_0<0$. Therefore the current case does not correspond to physical solutions for applications considered in this work.
\end{remark}

\section{Gravity-dependent flows}\label{sec:grav}

We now take gravity effects in the main mixture flow model \eqref{GalInv} into account: $g\neq 0$. In the dimensionless equations \eqref{ND}, let $\kappa \thicksim \epsilon^{\alpha +\beta}$, which may represent a gentle slope, or model a vertical flow of a bubbly fluid with a small $\kappa$ (see Remark \ref{SmallKappa}). We generalize the Gardner-Morikawa scale transformation to include a variable  reference speed $a(\tau)$:
\begin{equation}\label{eq:GM:gen}
  \xi = \epsilon^{\alpha}\left(x - a(\tau) t\right), \qquad \tau = \epsilon^{\alpha + \beta}t.
\end{equation}
The leading-order conditions arising from the substitution of the perturbation expansions \eqref{expansion} into the DEs \eqref{ND} in this case become
\begin{subequations}\label{eq:loword:vert}
\begin{equation}
\pder{}{\xi}u^{(1)} = \frac{1}{\rho_0}\left(\pder{a}{\tau} \tau + a\right)\pder{}{\xi}\rho^{(1)},
\end{equation}
\begin{equation}
P^{(1)} = P_2^{(1)},
\end{equation}
\begin{equation}
P_2^{(1)} = -\frac{3P_0}{R_0}R^{(1)}+\frac{3\kappa a \tau }{((R_0^3-X+1)R_0)}R^{(1)},
\end{equation}
\begin{equation}
R^{(1)} = -\frac{(R_0^3-X+1)^2}{3 R_0} \rho^{(1)},
\end{equation}
\begin{equation}
\pder{}{\xi}P^{(1)} = \rho_0 \left(\pder{a}{\tau} \tau + a\right) \pder{}{\xi}u^{(1)},
\end{equation}
\end{subequations}
holding for an arbitrary set of positive constants $\alpha$, $\beta$, $q$ and $k$. The relations \eqref{eq:loword:vert} equations contain a constraint on $a(\tau)$; it is satisfied by
\begin{equation}
\barr
a(\tau) =&-\dfrac{P_0 R_0^6-2 P_0 R_0^3X-R_0^3C_1^2+2P_0 R_0^3+P_0 X^2-2 P_0 X+P_0}{\kappa
\left(R_0^3 - X + 1\right)\tau}\\[2ex]
& -\dfrac{\kappa \rho_0(R_0^3-X+1)^2\tau}{4R_0^3}+C_1,
\earr
\end{equation}
where $C_1$ is an arbitrary constant. By taking $C_1 = a_0$ \eqref{a_0}, the expression for $a(\tau)$ simplifies to
\begin{equation}
a(\tau) = \frac{1}{4}\kappa R_0^{-3}(R_0^3-X+1)\tau  +R_0^{-3/2} P_0^{1/2}(R_0^3-X+1).
\end{equation}
In order to reduce to a single equation satisfied by one of the leading-order parameter perturbations, one follows the same process as outlined in Section \ref{pertAnalysis}. In particular, the following result can be obtained.

\begin{proposition}\label{prop:burg:vcoef}
For an arbitrary choice of positive exponents satisfying $\alpha = \beta = q = k$ in the generalized  Gardner-Morikawa transformation \eqref{eq:GM:gen}, the lowest-order $\epsilon$-dependent terms of the equations \eqref{ND} are satisfied by the first two terms of the asymptotic expansions \eqref{expansion}, with $\rho^{(1)}$ being an arbitrary solution of a variable-coefficient Burgers equation
\begin{equation}\label{eq:grav:vcb}
  A(\tau) \pder{\rho^{(1)}}{\tau} =  B(\tau) \pderr{\rho^{(1)}}{\xi} +C(\tau)\rho^{(1)} \pder{\rho^{(1)}}{\xi} + D(\xi) \rho^{(1)}_{\xi} + E\rho^{(1)},
\end{equation}
and $P^{(1)}$, $P_2^{(1)}$, $R^{(1)}$, $u^{(1)}$ determined from \eqref{eq:loword:vert}, where $A$, $B$, and  $C$ are certain polynomials in terms of $\tau$, $D$ is a polynomial in terms of $\xi$, and $E$ is a constant.
\end{proposition}

Thus any solution of the variable-coefficient Burgers equation \eqref{eq:grav:vcb} yields an approximate solution of the full model \eqref{GalInv} with $g\neq 0$; this solution is given by the $\mathcal{O}(1)$ and $\mathcal{O}(\epsilon^q)$ first two terms of the expansion \eqref{expansion}.

As in Section \ref{pertAnalysis}, the $\mathcal{O}(\epsilon)$ terms of the expansions \eqref{expansion} of other state variables are related with $\rho^{(1)}$. The expressions for these terms, as well as $A$, $B$, $C$, $D$, and $E$, are not written here explicitly for the sake of brevity, and are readily derived by an eager reader.

\section{The case of nonzero surface tension} \label{sec:surfT}

When surface tension of the bubbles is not negligible, the dimensionless Rayleigh-Plesset equation \eqref{PfuncND} (with tildes omitted, as usual) is modified as follows \cite{church1995effects, doinikov2009modeling}:
\begin{equation}\label{PfuncND2}
P = P_2 - \left(R \matD_t^2R + \frac{3}{2}\left(\matD_tR\right)^2 + \frac{4 }{3 }\left(\frac{\matD_t R}{R}\right)+ \frac{2 \sigma' }{R}\right),
\end{equation}
where
\begin{equation}
  \sigma' = \frac{\rho_1\,\sigma}{\left(\frac{4}{3}\pi N \rho_1\right)^{1/3} \mu^2}
\end{equation}
is a constant dimensionless tension parameter, and $\sigma$ is the coefficient of surface tension of the bubble surface.

The full three-dimensional model of the bubbly flow is consequently given by the equations \eqref{GalInv3D:conti}, \eqref{GalInv3D:mom}, \eqref{PfuncND2}, \eqref{GalInv3D:P2}, and \eqref{GalInv3D:Rrho}.

\subsection{Exact equilibrium and traveling wave solutions}

Similarly to  Section \ref{equilibriumSection}, we can derive an equilibrium solution of the extended with non-zero surface tension. Seeking equilibrium solutions in one space dimension, in the form \eqref{afs:equil:dimless:1}, we find that the system \eqref{ND} (with \eqref{PfuncND} replaced by \eqref{PfuncND2}) admits equilibrium solutions satisfying the previously derived equations \eqref{afs:equil:dimless:u},
where $u_e(x)$ satisfies an amended version of the ODE \eqref{prop:eqsol:1:sol}:
\begin{equation}\label{prop:eqsol:1:sol:surften}
\barr
P_e(x) = P_{2e}(x) - u_e(x)\left(R_e(x)\der{^2 R_e(x)}{x^2} + \der{u_e(x)}{x} R_e(x)\der{R_e(x)}{x} \right.\\[2ex]
 \qquad \qquad +\left.\dfrac{3}{2}\left(\der{R_e(x)}{x}\right)^2 + \dfrac{4}{3 R_e(x)} \der{R_e(x)}{x} + \dfrac{2 \sigma'}{R_e(x)}\right).
\earr
\end{equation}

In the case of $u_e(x) = 0$, the following equilibrium solutions arise:
\begin{equation} \label{afs:equil:dimless:surf}
\barr
u_e(x) = 0,\quad
P_e(x) = P_0 + \displaystyle\int\kappa \rho_e(x) d x,   \\[2ex]
P_{2e}(x) = P_e(x) + \dfrac{2 \sigma'}{R_e(x)}, \quad
R^3_e(x)=\dfrac{1}{\rho_e(x)}+X-1,
\earr
\end{equation}
where $\rho_e(x)$ is an arbitrary initial density distribution.

\medskip
Similarly to the zero-tension case (Section \ref{sec:trv}), the equilibrium solutions \eqref{afs:equil:dimless:surf} and the Galilei invariance can be used to generate time-dependent exact solutions of the form \eqref{eq:galil:sols} of the model \eqref{ND}, \eqref{PfuncND2}.

\subsection{Generalized power series solutions}

In order to derive series-type and approximate solutions of the extended model with nonzero surface tension, one can again following the procedure of Section \ref{pertAnalysis} to analyze small perturbations of, for example, horizontal ($\kappa = g = 0$) equilibrium flows with nonzero surface tension.

Perturbing the equations \eqref{ND} (with \eqref{PfuncND} replaced by \eqref{PfuncND2}) around the equilibrium \eqref{afs:equil:dimless:surf} where we choose $\rho_e(x)=\rho_0=\const$, $P_e(x)=P_{2e}(x)=\const$, $R(x)=R_0=\const$, we pose Su-Gardner-type generalized power series \eqref{expansion} and the Gardner-Morikawa coordinate transformation  \eqref{longWave} sharing the small parameter $\epsilon$. We find that the equations are satisfied to their lowest-order terms in  $\epsilon$ when the $\mathcal{O}(\epsilon^q)$ terms of the solution series \eqref{expansion} satisfy
\begin{subequations} \label{lowestOrdst}
\begin{equation}
\pder{}{\xi}u^{(1)} = \frac{a_0}{\rho_0}\pder{}{\xi}\rho^{(1)},
\end{equation}
\begin{equation}
P^{(1)} = P_2^{(1)} + \frac{2 \sigma'}{R_0^2} R^{(1)},
\end{equation}
\begin{equation}
P_2^{(1)} = -\frac{3(P_0 R_0 + 2 \sigma')}{R_0}R^{(1)},
\end{equation}
\begin{equation}
R^{(1)} = -\frac{(R_0^3-X+1)^2}{3 R_0} \rho^{(1)},
\end{equation}
\begin{equation}
\pder{}{\xi}P^{(1)} = \rho_0 a_0 \pder{}{\xi}u^{(1)}.
\end{equation}
\end{subequations}
These equations involve a constraint on $a_0$ given by
\begin{equation}\label{a0:tens}
  a_0 =  \frac{(9R_0 P_0 + 12 \sigma')^{1/2}(R_0^3-X+1)}{3 R_0^2},
\end{equation}
which is an obvious generalization of \eqref{a_0} onto the case $\sigma'\ne 0$.

It can be shown that perturbations arise from solutions of a single PDE in the following specific cases.

\medskip\noindent\textbf{A) The Burgers case.} First, taking the familiar case with $\alpha = \beta = q = k$, we again find that the next order of $\epsilon$ in the bubbly fluid model is satisfied when $\rho^{(1)}$ satisfies the Burgers equation:
\begin{equation}
  \pder{\rho^{(1)}}{\tau} = A \pderr{\rho^{(1)}}{\xi} + B\rho^{(1)}\pder{\rho^{(1)}}{\xi},
\end{equation}
where $A$ and $B$ are constants given by
\[
     A =\dfrac { \left(P_0^{2}R_0^{5} + R_0^{4}\left( 4P_0\sigma'+ \tfrac{13}{3}T_0 \right)  +4R_0^{3}{
\sigma'}^{2} + [(P_0R_0+2\sigma')^2 +\tfrac{4}{3}R_0T_0]\left(1- X\right) \right)\left( R_0^{3
}-X+1 \right) }{6R_0^{4}T_0},
\]
\[
      B =-{3^{1/2}\dfrac { \left( R_0^{3}-X+1 \right) ^{3}}{(3\,
P_0\,R_0+4\sigma')^{1/2} R_0^{5}} \left( P_0R_0
+{\dfrac {14}{9}\sigma'} \right) }\,.
\]

\medskip\noindent\textbf{B) The linear diffusion equation case.} For the choice $\alpha = \beta = q/2 = k$, one again arrives at the diffusion equation for the lowest-order perturbation:
\begin{equation}
\pder{\rho^{(1)}}{\tau} = A\pderr{\rho^{(1)}}{\xi},
\end{equation}
where $A$ is a constant given by
\[
  A= \dfrac { \left( P_0^{2}R_0^{5}+ R_0^{4}\left( 4P_0\sigma'+\tfrac{13}{3}T_0 \right) +4R_0^{3}{
\sigma'}^{2} + \left[(P_0R_0+2{\sigma'})^{2} + \tfrac{4}{3}R_0T_0\right](1-X)\right)  \left( R_0^{3
}-X+1 \right)}{6R_0^{4}T_0}.
\]
\medskip\noindent\textbf{C) The Korteweg-de Vries equation case.} When the constant exponents are related by $\alpha = \beta = q/2 = k/2$, similarly to the result in Section \ref{sec:pert:KdV}, one again observes that  the lowest-order perturbation satisfies the Korteweg-de Vries equation
\begin{equation}
\pder{\rho^{(1)}}{\tau} = A \pder{^3\rho^{(1)}}{\xi} + B\rho^{(1)}\pder{\rho^{(1)}}{\xi}
\end{equation}
with tension-dependent constant parameters $A$ and $B$ given by
\[
\barr
 A =& Q \Big[{ \left( P_0 R_0^
{8} \left( \gamma-1 \right)+2\sigma' R_0^{7}\left( \gamma-1 \right) -{\tfrac {169}{27}}R_0^{6} + {\tfrac{104}{27}}R_0^{3}\left(X - 1\right) \right)}  \\
& + \left(2P_0 R_0^{5}  -4\sigma' R_0^{4}\right) \left( \gamma-1 \right)  \left(1-X\right) \\
& +\left(P_0  \left(\gamma-1 \right) ^{2}+2\sigma' R_0
 \left( \gamma-1 \right)-{\tfrac{16}{27}} \right)\left( 1-X \right)^{2}\Big]\,,
\earr
\]
\[
\barr
B &=-3^{1/2}{\dfrac { \left( R_0^{3}-X+1 \right) ^{3}}{R_0^{5} (3
P_0R_0+4\sigma')^{1/2}} \left( P_0R_0
+{\tfrac {14}{9}\sigma'} \right) }\, ,
\earr
\]
where we have denoted
\[
Q=-\dfrac{\left( P_0\,R_0+\tfrac{4}{3}\sigma' \right)  \left( R_0^{3}-X+1 \right)}{2R_0^{4} (9P_0R_0+12\sigma')^{1/2} \left( \gamma-1 \right)  \left( P_0R_0+2
\sigma' \right) }\,.
\]

Again, similarly to the result in Section \ref{sec:pert:KdV}, the Korteweg-de Vries equation holds when the initial values of the physical fields satisfy a special condition, here given by
\begin{equation}\label{kdvconstraint2}
  T_0 =-3\dfrac {P_0^{2}R_0^{5}+4R_0^{4}P_0\sigma' + (P_0R_0+2\sigma')^2 (1-X)+4\,R_0^{3}{\sigma'
}^{2}}{R_0 \left( 13R_0^{3}-4X+4 \right) }.
\end{equation}
Again, since physically, $P_0, R_0>0$ and $0<X<1$, it turns out that the reference initial temperature $T_0$ must be negative, which does not correspond to a physical reality for applications considered in this work.


%
%

\section{Discussion}  \label{sec:Conclusion}

In this paper, an extended Galilei-invariant physical model \eqref{GalInv3D} of a three-dimensional flow of a multiphase continuum, namely, a viscous fluid containing bubbles, was presented. The model was based on both physics of a single bubble and the mixture flow as a whole; it incorporates multiple physical effects including gravity, viscosity, and bubble surface tension. Two dimensionless versions of the model, \eqref{ND} and \eqref{physND}, were derived. In particular, in the general non-dimensionalization \eqref{ND} leads to a parameter reduction, i.e., it involves a fewer number of parameters than the dimensional model \eqref{GalInv}, and these parameters are dimensionless.
The ``physical" non-dimensionalization \eqref{physND} of the model \eqref{GalInv3D}, \eqref{GalInv} is based on characteristic values of the flow parameters. It leads to a set of dimensionless equations that include the classical fluid dynamics constants, such as the  Reynolds number $\Re$, the Euler  number $\Eu$, the bubble size/characteristic length ratio $\delta$, the thermal constant $W$, the dimensionless typical gas content $\B$, and the gravity constant $\varkappa$. The importance of this non-dimensional version of the model lies in the fact that for any bubbly flow regime of interest, one can compute $\Re$, $\Eu$, $\delta$, $W$, $\B$, and $\varkappa$, and determine the relative magnitudes of coefficients at different terms in the governing equations (see Table \ref{tab:1} where the parameter values for three sample flow types are presented). On the contrary, in the general non-dimensionalization \eqref{ND}, all terms in the equations would generally have similar orders.

Exact and approximate solutions of the one-dimensional reductions \eqref{GalInv} of the full bubble-fluid mixture model \eqref{GalInv3D} were analyzed, in the cases of absent and present gravity terms. In particular, for the dimensionless equations \eqref{ND}, equilibrium solutions holding for an arbitrary initial velocity distribution, as well as static equilibrium solutions and traveling wave solutions, were considered in Sections \ref{equilibriumSection} and \ref{sec:trv}. In particular, through Galilei transformations \eqref{eq:GalTr:newsol}, every equilibrium solution can be mapped into time-dependent traveling wave solutions \eqref{eq:galil:sols} moving with an arbitrary constant speed $c$ (Proposition \eqref{cololl:galil}).

In order to systematically construct approximate solutions of the one-dimensional  mixture model \eqref{GalInv} in the case of a vanishing gravity term, in Section \ref{pertAnalysis}, generalized asymptotic expansions of solutions of the bubbly fluid model in terms of a small parameter $\epsilon$, about the constant equilibrium state, were considered, in terms of the Gardner-Morikawa variables \eqref{longWave} (large wavelength, slow time) involving the same small parameter. It was shown that for several choices of Gardner-Morikawa exponents, leading terms of the solution perturbation series arise from solutions of single PDEs, namely, the Burgers equation \eqref{BurgersEq}, the linear diffusion equation \eqref{eq:rho:heat}, or the Korteweg-de Vries equation \eqref{eq:KdV1}. Thus exact solutions of these three classical models can be used to construct closed-form approximate solutions of the bubbly fluid model \eqref{GalInv}. This is illustrated for the classical $tanh$ kink-type traveling wave solution of the Burgers equation \eqref{BurgSol} in Section \ref{sec:ND:Burg}. The quality of approximation is demonstrated through a comparison with a numerical solution based on the method of lines (Appendix \ref{Sec:Num}).

Flows that essentially involve the gravity were considered in Section \ref{sec:grav}; there, in order to reduce the problem to a single nonlinear PDE, the Gardner-Morikawa scale transformations needed to be generalized (formula \eqref{eq:GM:gen}) to include a variable wave speed $a(\tau)$. It is shown that this choice can lead, for example, to a variable-coefficient Burgers equation \eqref{eq:grav:vcb} satisfied by the leading-term density perturbation. Consequently, any solution of a single PDE \eqref{eq:grav:vcb} yields, through \eqref{eq:loword:vert}, yields an approximate two-term solution (first two terms of \eqref{expansion}) $\rho(x,t)$, $u(x,t)$, $P(x,t)$, $P_2(x,t)$, $R(x,t)$ of the bubbly fluid model \eqref{ND} with a nonzero gravity term: $g, \kappa\ne 0$.

The bubble-liquid mixture flow model \eqref{GalInv3D} was further extended in Section \ref{sec:surfT} by incorporating surface tension effects into the (modified) Rayleigh-Plesset equation \eqref{PfuncND2}. Equilibrium and traveling-wave one-dimensional solutions were constructed, and generalized power series solutions were considered. It was shown that leading-order terms of the solution perturbations around the static equilibria for such flows can also satisfy a single simplified PDE, such as linear diffusion, Burgers, or Korteweg-de Vries equation, whose coefficients now essentially involve tension. Hence again solutions of a single PDE give rise to approximate solutions of the full one-dimensional flow model \eqref{GalInv}.


The presented model provides a significant extension and modification of the one obtained in a recent work \cite{Kud_Sine,kudryashov2014extended}. The differences include the full three-dimensional formulation, the Galilean invariance, the use of the full Navier-Stokes instead of Euler mixture flow equations, nonzero gravity, and a different physical assumption on the heat transfer term between the two phases.


An interesting aspect of Kudryashov and Sinelshchikov \cite{kudryashov2014extended} is their use of asymptotic expansions to derive a new third and fourth-order equation for small solution perturbations. We were able to obtain these equations up to first order in $\epsilon$ by taking similar choices of parameters as Kudryashov and Sinelshchikov \cite{kudryashov2014extended} ($\beta = q = k = 1$, and $\alpha = m $). In this case, leading terms of the solution perturbation series arise from the following non-linear equation
\begin{equation} \label{KdV:Burg}
\pder{}{\tau} \rho^{(1)} + C_1 \rho^{(1)}\pder{\rho^{(1)}}{\xi} + C_2\epsilon^{m - 1}\pder{^2\rho^{(1)}}{\xi^2} + C_3 \epsilon^{2m - 1}\pder{^3\rho^{(1)}}{\xi^3} = 0,
\end{equation}
where $C_1$, $C_2$, and $C_3$ are constants. Depending on the choice of $m$, \eqref{KdV:Burg} collapses into either the Burgers' case in Section  \ref{sec:ND:Burg} for $m =1$ or Korteweg-de Vries case in Section  \ref{sec:pert:KdV} for $m = 1/2$ including the requirement on $C_2$ vanishing. It was not feasible to obtain the higher-order terms due to the time derivatives of gas pressure in our heat transfer equation, which prevents splitting of the orders of $\epsilon$. Our results show agreement with the work of Kanagawa \emph{et al} \cite{kanagawa2010unified}, in the case of expansion parameters $\beta = q = k = 1$, and $\alpha = m$, which corresponds to their case of the high frequency and long wavelength band-up. This agreement is up to the constants of the resulting PDE; the difference is a result of different choice of bubble surface equations.

%
%

The physical model presented in this work can be extended further. Physically relevant extensions would include the consideration of heat loss to the environment, a description of bubble formulation and collapse, and accounting for the gas exchange between the bubbles and the surrounding fluid; the latter is important for a more accurate description of the mixture dynamics in the magma chamber as well as the vertical volcanic conduit in a wider range of depths.

\medskip  The generalized asymptotic series analysis of the bubbly fluid models considered in this paper, as well as in the related literature, can also potentially be extended, to systematically seek additional classes of Gardner-Morikawa exponents $\alpha$, $\beta$, $q$, $k$ that would lead to other non-constant values of first-order solution perturbations, and perhaps to the discovery of new simplified PDEs that describe the perturbation dynamics. An example of such work would be the nonlinear Schr\"{o}dinger that has been derived in \cite{kanagawa2010unified} from similar equations. A related avenue that could lead to new important results is the use of physical non-dimensionalizations like \eqref{physND}, where equations would already involve small parameters coming directly from the physical setting (e.g, Table \ref{tab:1}). Then the choice of the powers of the small parameter $\epsilon$ in the Su-Gardner-Morikawa-type expansions \eqref{expansion}, \eqref{longWave} naturally can, and should, be informed by the different scales of physical parameters contained in the dimensionless model equations.

\section*{Acknowledgements}
The authors are grateful to Jaden Dasiuk for help at the initial stages of this work, and to NSERC of Canada for the financial support through USRA and Discovery grants.

{\footnotesize
\bibliography{bubble3}

\begin{thebibliography}{10}

\bibitem{nigmatulin1990dynamics}
R.~I. Nigmatulin, {\em Dynamics of Multiphase Media}, vol.~2.
\newblock CRC Press, 1990.

\bibitem{gidaspow1994multiphase}
D.~Gidaspow, {\em Multiphase Flow and Fluidization: Continuum and Kinetic
  Theory Descriptions}.
\newblock Academic press, 1994.

\bibitem{kolev2005multiphase}
N.~I. Kolev and N.~Kolev, {\em Multiphase Flow Dynamics}, vol.~1.
\newblock Springer, 2005.

\bibitem{brennen2005fundamentals}
C.~E. Brennen, {\em Fundamentals of Multiphase Flow}.
\newblock Cambridge University Press, 2005.

\bibitem{passman1984theory}
S.~L. Passman, J.~W. Nunziato, and E.~K. Walsh, ``A theory of multiphase
  mixtures,'' in {\em Rational Thermodynamics}, pp.~286--325, Springer, 1984.

\bibitem{melnik}
O.~Melnik, A.~Barmin, and R.~Sparks, ``Dynamics of magma flow inside volcanic
  conduits with bubble overpressure buildup and gas loss through permeable
  magma,'' {\em Journal of Volcanology and Geothermal Research}, vol.~143,
  pp.~53--68, 2005.

\bibitem{Kud_Sine}
N.~Kudryashov and D.~Sinelshchikov, ``Nonlinear waves in bubbly liquids with
  considersation for viscosity and heat transfer,'' {\em Physics Letters A},
  vol.~374, pp.~2011--2016, 2010.

\bibitem{wang2011thermodynamic}
Y.~Wang and M.~Oberlack, ``A thermodynamic model of multiphase flows with
  moving interfaces and contact line,'' {\em Continuum Mechanics and
  Thermodynamics}, vol.~23, no.~5, pp.~409--433, 2011.

\bibitem{kallendorf2012conservation}
C.~Kallendorf, A.~F. Cheviakov, M.~Oberlack, and Y.~Wang, ``Conservation laws
  of surfactant transport equations,'' {\em Physics of Fluids}, vol.~24,
  no.~10, p.~102105, 2012.

\bibitem{cheviakov2017symbolic}
A.~F. Cheviakov, ``Symbolic computation of equivalence transformations and
  parameter reduction for nonlinear physical models,'' {\em Computer Physics
  Communications}, vol.~220, pp.~56--73, 2017.

\bibitem{savage1979gravity}
S.~B. Savage, ``Gravity flow of cohesionless granular materials in chutes and
  channels,'' {\em Journal of Fluid Mechanics}, vol.~92, no.~1, pp.~53--96,
  1979.

\bibitem{hon1994emplacement}
K.~Hon, J.~Kauahikaua, R.~Denlinger, and K.~Mackay, ``Emplacement and inflation
  of pahoehoe sheet flows: Observations and measurements of active lava flows
  on {K}ilauea {V}olcano, {H}awaii,'' {\em Geological Society of America
  Bulletin}, vol.~106, no.~3, pp.~351--370, 1994.

\bibitem{JGRB:JGRB11445}
L.~Keszthelyi and S.~Self, ``Some physical requirements for the emplacement of
  long basaltic lava flows,'' {\em Journal of Geophysical Research: Solid
  Earth}, vol.~103, no.~B11, pp.~27447--27464, 1998.

\bibitem{sakimoto1998flow}
S.~Sakimoto and M.~Zuber, ``Flow and convective cooling in lava tubes,'' {\em
  Journal of Geophysical Research: Solid Earth}, vol.~103, no.~B11,
  pp.~27465--27487, 1998.

\bibitem{park1984dynamics}
S.~Park and J.~D. Iversen, ``Dynamics of lava flow: Thickness growth
  characteristics of steady two-dimensional flow,'' {\em Geophysical Research
  Letters}, vol.~11, no.~7, pp.~641--644, 1984.

\bibitem{rayleigh1917viii}
L.~Rayleigh, ``{VIII}. on the pressure developed in a liquid during the
  collapse of a spherical cavity,'' {\em The London, Edinburgh, and Dublin
  Philosophical Magazine and Journal of Science}, vol.~34, no.~200, pp.~94--98,
  1917.

\bibitem{plesset1954growth}
M.~Plesset and S.~A. Zwick, ``The growth of vapor bubbles in superheated
  liquids,'' {\em Journal of Applied Physics}, vol.~25, no.~4, pp.~493--500,
  1954.

\bibitem{foldy1945multiple}
L.~L. Foldy, ``The multiple scattering of waves. {I}. {G}eneral theory of
  isotropic scattering by randomly distributed scatterers,'' {\em Physical
  Review}, vol.~67, no.~3-4, p.~107, 1945.

\bibitem{wijngaarden1972one}
L.~van Wijngaarden, ``One-dimensional flow of liquids containing small gas
  bubbles,'' {\em Annual Review of Fluid Mechanics}, vol.~4, pp.~369--396,
  1972.

\bibitem{miksis1991effective}
M.~J. Miksis and L.~Ting, ``Effective equations for multiphase flows -- waves
  in a bubbly liquid,'' {\em Advances in Applied Mechanics}, vol.~28,
  pp.~141--260, 1991.

\bibitem{jordan2004propagation}
P.~Jordan and C.~Feuillade, ``On the propagation of harmonic acoustic waves in
  bubbly liquids,'' {\em International Journal of Engineering Science},
  vol.~42, no.~11, pp.~1119--1128, 2004.

\bibitem{jordan2006propagation}
P.~Jordan and C.~Feuillade, ``On the propagation of transient acoustic waves in
  isothermal bubbly liquids,'' {\em Physics Letters A}, vol.~350, no.~1,
  pp.~56--62, 2006.

\bibitem{becher2012handbook}
H.~Becher and P.~N. Burns, {\em Handbook of Contrast Echocardiography: Left
  Ventricular Function and Myocardial Perfusion}.
\newblock Springer Science \& Business Media, 2012.

\bibitem{goldberg1994ultrasound}
B.~B. Goldberg, J.-B. Liu, and F.~Forsberg, ``Ultrasound contrast agents: a
  review,'' {\em Ultrasound in Medicine and Biology}, vol.~20, no.~4,
  pp.~319--333, 1994.

\bibitem{szabo2004diagnostic}
T.~L. Szabo, {\em Diagnostic Ultrasound Imaging: Inside Out}.
\newblock Academic Press, 2004.

\bibitem{kanagawa2015two}
T.~Kanagawa, ``Two types of nonlinear wave equations for diffractive beams in
  bubbly liquids with nonuniform bubble number density,'' {\em The Journal of
  the Acoustical Society of America}, vol.~137, no.~5, pp.~2642--2654, 2015.

\bibitem{kanagawa2011nonlinear}
T.~Kanagawa, T.~Yano, M.~Watanabe, and S.~Fujikawa, ``Nonlinear wave equation
  for ultrasound beam in nonuniform bubbly liquids,'' {\em Journal of Fluid
  Science and Technology}, vol.~6, no.~2, pp.~279--290, 2011.

\bibitem{Nakoryakov1972}
V.~E. Nakoryakov, V.~V. Sobolev, and I.~R. Shreiber, ``Longwave perturbations
  in a gas-liquid mixture,'' {\em Fluid Dynamics}, vol.~7, pp.~763--768, Sep
  1972.

\bibitem{kudryashov2014extended}
N.~A. Kudryashov and D.~I. Sinelshchikov, ``Extended models of non-linear waves
  in liquid with gas bubbles,'' {\em International Journal of Non-Linear
  Mechanics}, vol.~63, pp.~31--38, 2014.

\bibitem{kanagawa2010unified}
T.~Kanagawa, T.~Yano, M.~Watanabe, and S.~Fujikawa, ``Unified theory based on
  parameter scaling for derivation of nonlinear wave equations in bubbly
  liquids,'' {\em Journal of Fluid Science and Technology}, vol.~5, no.~3,
  pp.~351--369, 2010.

\bibitem{nakoryakov1993wave}
V.~Nakoryakov, B.~Pokusaev, and I.~Shreiber, {\em Wave Propagation in
  Gas-Liquid Media}.
\newblock CRC Press, 1993.

\bibitem{leighton2007derivation}
T.~Leighton, ``Derivation of the rayleigh-plesset equation in terms of
  volume,'' 2007.

\bibitem{brennen2013cavitation}
C.~E. Brennen, {\em Cavitation and bubble dynamics}.
\newblock Cambridge University Press, 2013.

\bibitem{stolper1980melt}
E.~Stolper and D.~Walker, ``Melt density and the average composition of
  basalt,'' {\em Contributions to Mineralogy and Petrology}, vol.~74, no.~1,
  pp.~7--12, 1980.

\bibitem{su1969korteweg}
C.~Su and C.~Gardner, ``Korteweg-de {V}ries equation and generalizations.
  {III}. {D}erivation of the {K}orteweg-de {V}ries equation and {B}urgers
  equation,'' {\em Journal of Mathematical Physics}, vol.~10, no.~3,
  pp.~536--539, 1969.

\bibitem{kudryashov2006nonlinear}
N.~Kudryashov and I.~Chernyavskii, ``Nonlinear waves in fluid flow through a
  viscoelastic tube,'' {\em Fluid Dynamics}, vol.~41, no.~1, pp.~49--62, 2006.

\bibitem{church1995effects}
C.~C. Church, ``The effects of an elastic solid surface layer on the radial
  pulsations of gas bubbles,'' {\em The Journal of the Acoustical Society of
  America}, vol.~97, no.~3, pp.~1510--1521, 1995.

\bibitem{doinikov2009modeling}
A.~A. Doinikov, J.~F. Haac, and P.~A. Dayton, ``Modeling of nonlinear viscous
  stress in encapsulating shells of lipid-coated contrast agent microbubbles,''
  {\em Ultrasonics}, vol.~49, no.~2, pp.~269--275, 2009.

\end{thebibliography}
\bibliographystyle{ieeetr}
}

\begin{appendix}
\section{The Numerical Method}\label{Sec:Num}

Since the dimensionless bubbly fluid model equations \eqref{ND} is not given by evolution equations, in order to solve it numerically, a non-standard procedure must be used. (Throughout this section, tildes in the equations of \eqref{ND} and related formulas are omitted.) We employ a modified method of lines, as follows.

The first step is to solve for and exclude $P(x,t)$ and $R(x,t)$ using the algebraic equations \eqref{PfuncND} and \eqref{F2funcND}, and the time derivatives $R_t$, $R_{tt}$ using the differential consequences of the necessary equations from \eqref{ND}. A new PDE system is obtained:
\begin{subequations}\label{num:sys1}
\begin{equation}\label{num:1}
\pder{\rho}{t} = -\pder{}{x}(\rho u  ),
\end{equation}
\begin{equation}\label{num:2}
\pder{P_2}{t}= - \gamma \frac{P_2 }{\rho R^3}\pder{u}{x} - \pder{P_2}{x} u  - 3 T_0(\gamma -1)\left(\frac{P_2 R^2}{P_0 R_0^3} - \frac{1}{R}\right),
\end{equation}
\begin{equation} \label{eq3:mixed}
 \rho \pder{u}{t} + \rho u \pder{u}{x} - \pderr{u}{x} + \pder{A}{x} + \frac{1}{3\rho^2 R}\pder{\rho}{x}\frac{\partial^2 u}{\partial x \partial t} + \frac{1}{3 \rho R^2}\pder{R}{x}\frac{\partial^2 u}{\partial x \partial t} - \frac{1}{3 \rho R}\frac{\partial^3 u}{\partial x^2 \partial t}  = 0,
\end{equation}
\end{subequations}
where $A(x,t)$ is given by
\begin{equation*}
A(x,t) = P_2 - \frac{u }{3\rho R}\pderr{u}{x} - \frac{1}{3 \rho R}\left(\pder{u}{x}\right)^2 - \frac{4 }{9\rho R^3}\pder{u}{x} + \frac{1}{18 \rho^2 R^4}\left(\pder{u}{x}\right)^2.
\end{equation*}
The PDE system \eqref{num:sys1} is also not a set of evolutionary equations, due to the presence of the mixed derivative ${\partial^3 u}/{\partial x^2 \partial t}$ in the PDE \eqref{eq3:mixed}. However, together with appropriate initial and boundary conditions, the PDEs \eqref{num:sys1} describe the dynamics of the fields  $\rho(x,t)$, $u(x,t)$, and $P_2(x,t)$. When the values of $\rho(x,t)$, $u(x,t)$, and $P_2(x,t)$ are obtained, the grid functions $P(x,t)$ and $R(x,t)$ are computed from the algebraic equations \eqref{PfuncND} and \eqref{F2funcND}.

We choose homogeneous space steps $h=L/N$, and  approximate the spacial derivatives in \eqref{num:sys1} using central differences, to solve for the  time derivatives of the unknowns at the nodes. The first two equations \eqref{num:1} and \eqref{num:2} are already in the right form, whereas \eqref{eq3:mixed} yields the difference form
\begin{equation} \label{eq3:dis}
\begin{split}
 &\left( \frac{1}{12}\,{\frac {\rho_{{i+1}}-\rho_{{i-1}}}{{{h}^{2}\rho_{{i}}}^{2}R_{{i
}}}}+\frac{1}{12}\,{\frac {R_{{i+1}}-R_{{i-1}}}{{h}^{2}\rho_{{i}}{R_{{i}}}^{2}
}}-\frac{1}{3}\,{\frac {1}{{h}^{2}\rho_{{i}}R_{{i}}}} \right) {\frac
{\rm d}{{\rm d}t}}u_{{i+1}}\\
&+ \left(- \frac{1}{12}\,{\frac {
\rho_{{i+1}}-\rho_{{i-1}}}{{{h}^{2}\rho_{{i}}}^{2}R_{{i}}}}-\frac{1}{12}\,{
\frac {R_{{i+1}}-R_{{i-1}}}{{h}^{2}\rho_{{i}}{R_{{i}}}^{2}}}-\frac{1}{3}\,{
\frac {1}{{h}^{2}\rho_{{i}}R_{{i}}}} \right) {\frac {\rm d}{{\rm d}t}}
u_{{i-1}}  + \left( \rho_{{i}}+\frac{1}{3}\,{\frac {1}{{h}^{2}\rho_{{
i}}R_{{i}}}} \right) {\frac {\rm d}{{\rm d}t}}u_{{i}} \\
 &+\frac{1}{2}\,{\frac {\rho_{{i}}u_{{i}} \left( u_{{i+1}}-u_{{i-1}}
 \right) }{h}}+\frac{1}{2}\,{\frac {{\it A}_{{i+1}}-{\it A}_{{i-1}}}{h}}-{
\frac {u_{{i+1}}-2\,u_{{i}}+u_{{i-1}}}{{h}^{2}}}=0.
\end{split}
\end{equation}
The approximate state variable values ($u_i$,  $\rho_i$,  $P_i$,  $(P_2)_i$,  $R_i$, $A_i$) at the $i^{th}$ node are functions of $t$. Denoting
\begin{equation*}
D_1(i) = \rho_{{i}}+\frac{1}{3}\,{\frac {1}{{h}^{2}\rho_{{
i}}R_{{i}}}},
\end{equation*}
\begin{equation*}
D_2(i) =  \frac{1}{12}\,{\frac {\rho_{{i+1}}-\rho_{{i-1}}}{{{h}^{2}\rho_{{i}}}^{2}R_{{i
}}}}+\frac{1}{12}\,{\frac {R_{{i+1}}-R_{{i-1}}}{{h}^{2}\rho_{{i}}{R_{{i}}}^{2}
}}-\frac{1}{3}\,{\frac {1}{{h}^{2}\rho_{{i}}R_{{i}}}},
\end{equation*}
\begin{equation*}
D_3(i) = - \frac{1}{12}\,{\frac {
\rho_{{i+1}}-\rho_{{i-1}}}{{{h}^{2}\rho_{{i}}}^{2}R_{{i}}}}-\frac{1}{12}\,{
\frac {R_{{i+1}}-R_{{i-1}}}{{h}^{2}\rho_{{i}}{R_{{i}}}^{2}}}-\frac{1}{3}\,{
\frac {1}{{h}^{2}\rho_{{i}}R_{{i}}}},
\end{equation*}
\begin{equation*}
B(i) = -\frac{1}{2}\,{\frac {\rho_{{i}}u_{{i}} \left( u_{{i+1}}-u_{{i-1}}
 \right) }{h}}-\frac{1}{2}\,{\frac {{\it A}_{{i+1}}-{\it A}_{{i-1}}}{h}}+{
\frac {u_{{i+1}}-2\,u_{{i}}+u_{{i-1}}}{{h}^{2}}},
\end{equation*}
one can rewrite the differential-difference equation \eqref{eq3:dis} as the following linear system:
\begin{equation}\label{eq:linsys:Ut}
\begin{pmatrix}
D_1(1) &D_2(1)& 0     & \cdots &  0     & D_3(1)   \\
D_3(2) & D_1(2) & D_2(2)  & 0      &\cdots  &   0      \\
 0 	   & D_3(3) &\ddots & \ddots &  0      & \vdots   \\
\vdots & 0    &\ddots &          &D_2(N-2)&   0      \\
 0     &\vdots&  0    &D_3(N-1)&D_1(N-1)& D_2(N-1) \\
D_2(N) & 0    & \cdots&   0    &D_3(N)  & D_1(N)
\end{pmatrix}
\begin{pmatrix}
u_t(1)\\
u_t(2) \\
\vdots\\
\\
\\
\vdots \\
u_t(N)
\end{pmatrix}
=
\begin{pmatrix}
B(1)\\
B(2) \\
\vdots\\
\\
\\
\vdots \\
B(N)
\end{pmatrix}.
\end{equation}
The equations \eqref{eq:linsys:Ut} can be solved for the time derivatives of $u$ at the nodes. We employed Matlab's \verb|ode23| to integrate in time the ODEs that arise from the solution of \eqref{eq:linsys:Ut} and the spatial discretizations of PDEs \eqref{num:1} and \eqref{num:2}.

Depending on the nature of the sought solution, relevant boundary conditions (for example, periodic ones with a spatial period $L$, or Dirichlet or Neumann boundary conditions at $x=0,~L$) and appropriate initial conditions are used.

\end{appendix}

\end{document}